\documentclass[%
 reprint,
 amsmath,amssymb,
 aps,
]{revtex4-2}

\usepackage{graphicx}
\usepackage{dcolumn}
\usepackage{bm}

\usepackage{amsthm}
\newtheorem{theorem}{Theorem}
\newtheorem{proposition}{Proposition}

\begin{document}

\preprint{APS/123-QED}

\title{Generation of Maximally Entangled States by Lyapunov Control Based on Entanglement Measure}

\author{Yunyan Lee}
\affiliation{%
School of Engineering,
Australian National University, Canberra ACT 2601, Australia
}%

\author{Ciann-Dong Yang}%
 \altaffiliation{cdyang@mail.ncku.edu.tw}
\affiliation{%
Department of Aeronautics and Astronautics, 
National Cheng Kung University, Taiwan
}%

\author{Daoyi Dong}%
\affiliation{%
School of Engineering,
Australian National University, Canberra ACT 2601, Australia
}%



\begin{abstract}
Maximally entangled states (MES) are highly valued in quantum information processing. In quantum control, the creation of MES is typically treated as a state transfer problem with a predefined MES as the target. However, this approach is limited by the requirement to predetermine the MES structure. This paper introduces an improved quantum Lyapunov control approach that relies on the quantum entanglement measure to construct the Lyapunov function, instead of using the distance between quantum states. This strategy enables the preparation of any MES, regardless of whether its structure is known beforehand, using a single control scheme. The proposed entanglement control technique is unaffected by the number of entangled subsystems since it targets the entanglement measure as a scalar. Initially applied to bipartite pure states, this method demonstrates its capability to generate Bell states and their equivalents. Subsequent applications to bipartite mixed states and multipartite systems illustrate that the technique can produce MES with unspecified structures.
\end{abstract}

\maketitle


\section{Introduction}
\label{sec:intro}
Quantum entanglement is an essential physical resource for quantum computation \cite{nielsen2010quantum} and quantum information processing (QIP) \cite{deng2017quantum}, such as quantum teleportation 
\cite{bennett1993teleporting, gordon2006generalized, hu2023progress}, quantum cryptography \cite{bennett1992quantum, pirandola2020advances}, superdense coding \cite{barreiro2008beating}. To complete the above QIP tasks, maximally entangled states (MES) are usually required. For example, the MESs of two-qubit systems are called Bell states, which are powerful resources for quantum communication \cite{horodecki_2009_quantum, Liu2021Two}. The MES of multiqubit systems, such as the GHZ state and W state \cite{horodecki_2009_quantum}, are the key ingredients of various quantum error correction codes and quantum communication protocols \cite{gour2007entanglement}. These requirements from QIP tasks drive research into the preparation and manipulation of MES. Additionally, a key advantage of MESs is their capacity to transform into any other state via local operations assisted by classical communication (LOCC) \cite{de2013maximally}. 


Definitions of MES vary based on entanglement measures, with bipartite pure states consistently having Bell states as MES, but multipartite states having varying MES depending on the measure used \cite{enriquez2016maximally}. For instance, among all three-qubit pure states the genuine three-party entanglement measured by three-tangle \cite{coffman2000distributed} has its maximum for the GHZ state, while the two-tangle $\tau_{2}$ and the persistence of entanglement \cite{briegel2001persistent} are largest for the W state.

Given the reliance of MES on specific entanglement measures and the incomplete understanding of many MES structures, the quantum control design aimed at generating MES should achieve improved performance based on the targeted entanglement measure compared to a specific quantum state. In response to this limitation, we propose an improved quantum Lyapunov control method that generates MES by maximizing the entanglement measure. This method is based on entanglement-dependent Lyapunov functions instead of distance-dependent Lyapunov functions.

Quantum Lyapunov control is a promising technique for quantum state transfer, utilizing feedback design for control field determination and open-loop application \cite{dong2023learning, Dong2022Quantum, Hou2012Optimal, kuang2008lyapunov,wang_2010_analysis}. This method has a wide range of applications, including driving quantum systems into decoherence-free subspaces \cite{yi2009driving, wang2010lyapunov}, accelerating adiabatic passage \cite{ran2017speeding}, implementing rapid Lyapunov control in finite-dimensional systems \cite{kuang2017rapid}, and achieving finite-time stabilization \cite{kuang2021finite}. Furthermore, Lyapunov-based feedback has been used to generate MES through open-loop control \cite{wang_2009_entanglement}. In the context of \(N\)-qubit systems, GHZ entanglement has been effectively produced \cite{liu2016lyapunov, Kuang2022Rapid}, with multipartite GHZ and W states generated using separate cavities and linear optical elements \cite{yu2007robust}. Additionally, Zou et al.  \cite{zou2003conditional} proposed a method for developing a GHZ state in four separate atoms housed in leaky cavities, employing linear optical elements.

In existing quantum Lyapunov control methods, the Lyapunov function $V(\rho) \geq 0$ serves as some measure of the distance between the current state $\rho$ and the target state $\rho_{d}$, and the Lyapunov control law is designed to make $V(\rho)$ decrease strictly with time until $V(\rho)=0$, at which the quantum state $\rho$ reaches the target state $\rho_{d}$. Especially, if we set $\rho_{d}$ equal to the Bell state, quantum Lyapunov control becomes a feasible solution for the preparation of MES. However, if the MES to be prepared has an unknown structure, as is often the case for bipartite mixed states and multipartite states, existing Lyapunov control methods are inadequate.

Different from the existing definition, the proposed Lyapunov function measures quantum state entanglement as $V(\rho)=\mathcal{N}-E(\rho) \geq 0$, where $E(\rho)$ is the desired entanglement measure, and $\mathcal{N}$ is the maximum of $E(\rho)$. The Lyapunov control law is to be designed so that $V(\rho)$ decreases (or equivalently, $E(\rho)$ increases) strictly with time until a steady state $\rho_{s s}$ is reached, at which $\dot{\rho}=0$ and $V\left(\rho_{s s}\right)=\mathcal{N}-$ $E\left(\rho_{s s}\right)=0$, that is, $E\left(\rho_{s s}\right)$ reaches its maximum $\mathcal{N}$ and the steady state $\rho_{s s}$ turns out to be the desired MES. In this control process, we do not need to specify a certain $\rho_{M E S}$ as the target state; instead, the control mechanism automatically generates the MES. The absence of a specified target state means that the steady-state $\rho_{ss}$ does not have to be the sole element in LaSalle's invariant set. Instead, LaSalle's invariant set comprises all the MESs to the measure $E(\rho)$.

Our study utilizes a Lyapunov function that depends on entanglement to create MES for bipartite pure states, bipartite mixed states, and multipartite states. For bipartite pure states, the effectiveness and accuracy of our method are confirmed by the known analytical form of MES. The Lyapunov control law based on a generalized measure of bipartite entanglement, steers the system towards LaSalle's invariant set, which includes Bell states and their equivalents. This approach automatically generates Bell states or their equivalents, eliminating the need for control field redesign for different Bell states, a requirement in existing methods. For bipartite mixed states, despite the absence of a method that prepares maximally entangled mixed states (MEMS) due to their unknown analytical form, our approach can produce MEMS with undefined forms. Regarding multipartite states, the form of MES is dictated by the chosen entanglement measure. As the number of subsystems involved in entanglement increases, the entanglement measurement becomes more complex. Nevertheless, our Lyapunov control strategy focused on the scalar multipartite entanglement measure rather than the states themselves remains effective. This allows the application of the same approach to the preparation of multipartite MES by constructing the Lyapunov function based on the multipartite entanglement measure.

Our approaches focus on the degree of entanglement, facilitating its extension to broader scenarios, as demonstrated through our discussions on bipartite pure states, bipartite mixed states, and multipartite states. Based on this idea, our methods, which focus on directly enhancing entanglement, could significantly contribute to advancements in quantum metrology. The results in \cite{long2022entanglement, wang2018entanglement} show the importance of entanglement in quantum metrology, particularly emphasizing its necessity for enhancing measurement precision. Enhanced entanglement levels within $N$-level quantum systems are crucial for parameter estimation accuracy \cite{toth2012multipartite, walborn2018quantum}. The connection between enhanced entanglement and improved measurement precision in quantum metrology demonstrates the potential applications of our proposed methods.

The remainder of this paper is organized as follows. After a brief mathematical preliminary given in Section \ref{sec: preliminaries}, Section \ref{sec: Lyapunov entangled function} introduces the Lyapunov entanglement function (LEF), serving as both a Lyapunov function for control law design and a quantitative entanglement index. Based on the LEF, Section \ref{sec: Lyapunov_design} details the design of Lyapunov control law for generating MES such that the controlled state converges to LaSalle's invariant set comprised of all the Bell states and their equivalent states. Section \ref{sec: numerical verification of maximum entanglement control} presents numerical demonstrations of generating MES using one control design with various initial states. Section \ref{sec: Lyapunov_design_mixed} extends the Lyapunov entanglement control to bipartite mixed states using the tilde decomposition method \cite{wootters_1998_entanglement}, verifying MEMS correctness with a specific mixed state class. Lastly, Section \ref{sec: Lyapunov_design_multi} considers two common multipartite entanglement measures, demonstrating successful multipartite MES generation with the proposed method.

\section{preliminaries}
\label{sec: preliminaries}
\subsection{Time evolution of pure and mixed states}
The time evolution of the pure state $|\psi\rangle$ of a closed quantum system satisfies the Schrödinger equation:
\begin{equation}
\label{eqn: 1}
i \hbar|\dot{\psi}(t)\rangle=\widehat{H}|\psi(t)\rangle, \quad \widehat{H}=\widehat{H}_{0}+\widehat{H}_{c}
\end{equation}
where $\widehat{H}_{0}$ is the internal Hamiltonian, and $\widehat{H}_{c}=\sum_{k=1}^{m} \widehat{H}_{k} u_{k}(t)$ is the time-dependent control Hamiltonian, which represents the interaction of the system with the control fields $u_{k}(t)$. The Hamiltonians $\widehat{H}_{0}$ and $\widehat{H}_{k}$ have to be Hermitian operators. Conveniently, we set $\hbar$ to $1$. It is easier to design the control field $u_{k}$ under the interaction picture defined by $\left|\psi_{I}(t)\right\rangle=$ $e^{i \widehat{H}_{0} \hbar t}|\psi(t)\rangle$, which satisfies
\begin{equation}
\label{eqn: 2}
i \hbar\left|\dot{\psi}_{I}(t)\right\rangle=\widehat{H}_{c, I}\left|\psi_{I}(t)\right\rangle, \widehat{H}_{c, I}=\sum_{k=1}^{m} \hat{A}_{k} u_{k}(t),
\end{equation}
where $\hat{A}_{k}$ is related to $\widehat{H}_{k}$ as $\hat{A}_{k}=e^{\mathrm{i} \hat{H}_{0} t / \hbar} \widehat{H}_{k}$. It can be shown that the expectation in the interaction picture is equal to the expectation in the Schrödinger picture.

When a quantum system is composed of multiple subsystems in different pure states, the system state becomes mixed and is described by the density operator
\begin{equation}
\label{eqn: 3}
\hat{\rho}=\sum p_{k}\left|\psi_{k}\right\rangle\left\langle\psi_{k}\right|=\sum p_{k} \hat{\rho}_{k} 
\end{equation}
where $p_{i}$ represents the weight of the component state $\left|\psi_{k}\right\rangle$ in the mixed state. For closed quantum systems, the time evolution of the density operator $\hat{\rho}$ satisfies the von Neumann equation,
\begin{equation}
\label{eqn: 4}
i \hbar \dot{\hat{\rho}}(t)=\widehat{H} \hat{\rho}(t)-\hat{\rho}(t) \widehat{H} \triangleq[\widehat{H}, \hat{\rho}(t)].
\end{equation}
The time evolution of $\hat{\rho}(t)$ described by \eqref{eqn: 4} is unitary by noting that $\hat{\rho}(t)$ can be expressed as $\hat{\rho}(t)=$ $\widehat{U}(t) \hat{\rho}(0) \widehat{U}^{\dagger}(t)$, where the unitary operator $\widehat{U}$ satisfies the following equation
\begin{equation}
\label{eqn: 5}
i \hbar \dot{U}(t)=\widehat{H} \widehat{U}(t), \widehat{U}(0)=I .
\end{equation}
The task of Lyapunov entanglement control amounts to finding the unitary operators $\widehat{U}$ to transform the initial state $\hat{\rho}(0)$ to $\rho_{\text{MES}}$ achieving maximum entanglement. The unitary transformation involved in $\rho_{\text{MES}}=\widehat{U} \hat{\rho}(0) \widehat{U}^{\dagger}$ is a global one that acts simultaneously on all the subsystems and is different from the unitary local transformation that only works on some of the subsystems and cannot increase the entanglement of $\hat{\rho}(0)$. Under the interaction picture, \eqref{eqn: 4} can be written as
\begin{equation}
\label{eqn: 6}
i \hbar \dot{\hat{\rho}}_{I}(t)=\left[\sum_{k=1}^{m} \hat{A}_{k} u_{k}(t), \hat{\rho}_{I}(t)\right] 
\end{equation}
with $\hat{A}_{k}$ defined in \eqref{eqn: 2} and $\hat{\rho}_{I}=e^{i \hat{H}_{0} t / \hbar} \rho e^{-i \hat{H}_{0} t / \hbar}$. Hereafter, we focus on the control of $\hat{\rho}_{I}$ and still denote $\hat{\rho}_{I}$ as $\hat{\rho}$.

\subsection{Matrix and vector representations}
The matrix representation of a quantum operator $\hat{A} \in \mathcal{H}$ is denoted by $A$, whose element is given by $A_{i j}=\left\langle e_{i}|\hat{A}| e_{j}\right\rangle$ with $\left|e_{i}\right\rangle$ being the basis of the Hilbert space $\mathcal{H}$. In this paper, we treat $A$ and $\hat{A}$ as equivalent expressions, such as $\hat{\rho} \triangleq \rho$ and $\widehat{H} \triangleq H$. Similarly, a quantum state $|v\rangle$ is equivalent to its vector representation $v$ with element given by $v_{i}=\left\langle e_{i} \mid v\right\rangle$, and its dual state $\langle v|$ is equivalent to $\left(v^{*}\right)^{T} \triangleq v^{\dagger}$. The Hilbert space dimension for a qubit is $d=2$, and we focus on qubit systems throughout the paper. Matrix $A$ (the associated operator $\hat{A}$ ) is said to be Hermitian, if $A=\left(A^{*}\right)^{T} \triangleq A^{\dagger}$, and to be skew-Hermitian, if $A^{\dagger}=-A$.

\subsection{Normal Matrix and its Spectral decomposition}
Matrix $A$ is said to be normal, if $A$ and $A^{\dagger}$ are commutative, i.e., $\left[A, A^{\dagger}\right]=A A^{\dagger}-A^{\dagger} A=0$. A normal matrix has a spectral decomposition $A=V \Lambda V^{\dagger}$, where $\Lambda$ is a diagonal matrix composed by the eigenvalues $\lambda_{k}$ of $A$, and $V=$ $\left[v_{1}, v_{2}, \cdots, v_{n}\right]$ is a unitary matrix containing the eigenvectors $v_{k}$ of $A$. When expressed by $v_{k}$, the spectral decomposition $A=V \Lambda V^{\dagger}$ becomes $A=\sum_{k=1}^{n} \lambda_{k} v_{k} v_{k}^{\dagger}$, which has an alternative expression in terms of the Dirac notation,
\begin{equation}
\label{eqn: 7}
A=\sum_{k=1}^{n} \lambda_{k}\left|\lambda_{k}\right\rangle\left\langle\lambda_{k}\right| ,
\end{equation}
where $\left|\lambda_{k}\right\rangle$ corresponds to the eigenvector $v_{k}$ and $\left\langle\lambda_{k}\right|$ to its conjugate transpose $v_{k}^{\dagger}$. Both Hermitian and skew-Hermitian matrices are normal and possess spectral decomposition, which provides a convenient way to evaluate the function of a normal matrix:
\begin{equation}
\label{eqn: 8}
f(A)=\sum_{k=1}^{n} f\left(\lambda_{k}\right)\left|\lambda_{k}\right\rangle\left\langle\lambda_{k}\right| .
\end{equation}
\subsection{Schmidt decomposition and partial trace}
A bipartite pure state described by $\left|\psi_{A B}\right\rangle \in \mathcal{H}_{A B}=\mathcal{H}_{A} \otimes$ $\mathcal{H}_{B}$ is said to be separable, if and only if it can be expressed as a tensor product of the states in the two subsystems:
\begin{equation}
\label{eqn: 9}
\left|\psi_{A B}\right\rangle=\left|\psi_{A}\right\rangle \otimes\left|\psi_{B}\right\rangle
\end{equation}
where $\left|\psi_{A}\right\rangle \in \mathcal{H}_{A}$ and $\left|\psi_{B}\right\rangle \in \mathcal{H}_{B}$. In terms of the orthogonal basis $\left|e_{k}^{A}\right\rangle \in \mathcal{H}_{A}$ and $\left|e_{k}^{B}\right\rangle \in \mathcal{H}_{B}$, any pure state $\left|\psi_{A B}\right\rangle$ has a Schmidt decomposition \cite{nielsen_1999_conditions} as
\begin{equation}
\label{eqn: 10}
\left|\psi_{A B}\right\rangle=\sum_{k=1}^{r} \sqrt{\alpha_{k}}\left|e_{k}^{A}\right\rangle \otimes\left|e_{k}^{B}\right\rangle ,
\end{equation}
where $r$ is the Schmidt rank of $\psi_{A B}$ and the Schmidt coefficient $\alpha_{k}>0$ is in decreasing order with $\sum_{k} \alpha_{k}=1$. A bipartite pure state $\psi_{A B}$ is separable, if and only if its Schmidt rank $r$ is equal to one, in which case \eqref{eqn: 10} reduces to \eqref{eqn: 9}. The density matrix of $\left|\psi_{A B}\right\rangle$ is $\rho_{A B}=\left|\psi_{A B}\right\rangle\left\langle\psi_{A B}\right|$ with $\left|\psi_{A B}\right\rangle$ given by \eqref{eqn: 10}. $\rho_{A}$ is called the reduced matrix of $\rho_{A B}$ obtained by taking the partial trace of $\rho_{A B}$ over the basis $\left|e_{k}^{B}\right\rangle$ of $\mathcal{H}_{B}$, i.e.,
\begin{equation}
\label{eqn: 11}
\rho_{A} \triangleq \operatorname{Tr}_{B}\left(\rho_{A B}\right)=\sum_{k=1}^{r}\left(I_{A} \otimes\left\langle e_{k}^{B}\right|\right) \rho_{A B}\left(I_{A} \otimes\left|e_{k}^{B}\right\rangle\right),
\end{equation}
where $I_{A}$ is the identity matrix in $\mathcal{H}_{A}$. In the following, we denote a reduced matrix as $\rho_M$.

\section{Lyapunov entangled function}
\label{sec: Lyapunov entangled function}
In this section, we propose a class of entanglement measures suitable for Lyapunov entanglement control by axiomatic approach \cite{plenio_2007_an, vidal_2000_entanglement}. A bipartite entanglement measure $E(\rho)$ is a mapping from density matrices into positive real numbers. Vidal \cite{vidal_2000_entanglement} characterized the entanglement measure $E(\rho)$ in terms of an entanglement monotone function $h$, which satisfies the following two properties.

\begin{enumerate}
  \item $h(\rho)$ is invariant under any unitary local transformation $U_{L}$, i.e., $h\left(U_{L} \rho U_{L}^{\dagger}\right)=h(\rho)$.
  \item $h(\rho)$ is concave downward, i.e., $h(\rho) \geq \lambda h\left(\rho_{1}\right)+(1-$ $\lambda) h\left(\rho_{2}\right)$ for $\rho=\lambda \rho_{1}+(1-\lambda) \rho_{2}, \lambda \in[0,1]$. 
\end{enumerate}
A class of entanglement measure satisfying the above property of entanglement monotone can be characterized explicitly as
\begin{equation}
\label{eqn: 12}
E_{G}(\rho)=G\left(\operatorname{Tr}\left(f\left(\rho_{M}\right)\right)\right), 
\end{equation}
where $\rho_{M}$ is the reduced matrix of $\rho$ and the trace operation of $f\left(\rho_{M}\right)$ makes $E_{G}(\rho)$ invariant under any unitary local transformation. The remaining properties of a qualified entanglement measure \(E_{G}(\rho)\) are ensured by applying appropriate conditions on the functions \(G\) and \(f\). These functions are required to be continuously twice differentiable when the density matrix \(\rho\) is non-separable, as will be derived in the following.

Firstly, we consider a bipartite pure state described by $\rho=$ $|\psi\rangle\langle\psi|$ and express its reduced matrix by the spectral decomposition:
\begin{equation}
\label{eqn: 13}
\rho_{M}=\lambda_{1}\left|\lambda_{1}\right\rangle\left\langle\lambda_{1}\left|+\lambda_{2}\right| \lambda_{2}\right\rangle\left\langle\lambda_{2}\right| .
\end{equation}
With the condition $\lambda_{1}+\lambda_{2}=1$, it is convenient to denote $\lambda_{1}=$ $\lambda$ and $\lambda_{2}=1-\lambda$ with $0 \leq \lambda \leq 1$ so that the general entanglement measure $E_{G}(\rho)$ in \eqref{eqn: 12} becomes a function of the eigenvalue $\lambda$ :
\begin{equation}
\label{eqn: 14}
E_{G}(\rho)=G(X(\lambda)), \quad X(\lambda)=\operatorname{Tr}\left(f\left(\rho_{M}\right)\right) .
\end{equation}
Here, when \( \lambda=0 \) or \( \lambda=1 \), it implies that the density matrix represents a separable state, i.e., \( E_{G}(\rho)=0 \). Using \eqref{eqn: 8} and \eqref{eqn: 14}, the function $X(\lambda)$ can be evaluated explicitly as
\begin{align}
\label{eqn: 15}
\begin{aligned}
    X(\lambda)&=\operatorname{Tr}\left(f\left(\rho_{M}\right)\right) \\
    &=\operatorname{Tr}\left(f\left(\lambda_{1}\right)\left|\lambda_{1}\right\rangle\left\langle\lambda_{1}\left|+f\left(\lambda_{2}\right)\right| \lambda_{2}\right\rangle\left\langle\lambda_{2}\right|\right) \\
& =f(\lambda)+f(1-\lambda),
\end{aligned}
\end{align}
where we note $\operatorname{Tr}\left(\left|\lambda_{k}\right\rangle\left\langle\lambda_{k}\right|\right)=\left\langle\lambda_{k} \mid \lambda_{k}\right\rangle=1$. As a result, we obtain a simple expression for the general entanglement measure $E_{G}(\rho)$ as
\begin{equation}
\label{eqn: 16}
E_{G}(\rho)=G(X(\lambda))=G(f(\lambda)+f(1-\lambda)) .
\end{equation}
Based on this concise expression, the required conditions on $G$ and $f$ to ensure $E_{G}(\rho)$ as a qualified entanglement measure can be derived straightforwardly as follows.

\begin{enumerate}
  \item $E_{G}(\lambda)=0$ for separable states. When the quantum state is separable, the rank of $\rho_{M}$ is 1, corresponding to $\lambda=0$ or $\lambda=1$. With \eqref{eqn: 16}, the requirement of $E_{G}(0)=E_{G}(1)=0$ turns out to be
    \begin{equation}
    \label{eqn: 17}
    G(f(0)+f(1))=0 .
    \end{equation}
  \item The positivity of $E_{G}(\lambda)$. $E_{G}(\lambda)$ must be positive for all entangled states, i.e., $E_{G}(\lambda)>0, \forall \lambda \in(0,1)$. This requirement is equivalent to
    \begin{equation}
    \label{eqn: 18}
    G(X)>0, \quad \forall X \neq f(0)+f(1) .
    \end{equation}
  \item $E_{G}^{\prime}(\lambda)=0$ as $\rho=\rho_{\text{MES}}$. When the quantum state $\rho$ is the MES, its reduced density matrix becomes \cite{Preskill_1999_Lecture}
  \begin{align}
  \label{eqn: 19}
    \left(\rho_{\text{MES}}\right)_{M}=\left[
    \begin{array}{ll}
1 / 2 & 0  \\
0 & 1 / 2
\end{array}\right],
\end{align}
i.e., $\lambda_{1}=\lambda_{2}=1 / 2$. This property requires that the derivative of $E_{G}(\lambda)$ must be zero at $\lambda=1 / 2$. This requirement is satisfied automatically by evaluating
\begin{align}
\label{eqn: 20}
\begin{aligned}
    E_{G}^{\prime}(\lambda)&=\frac{d G(X)}{d X} \frac{\partial X}{\partial \lambda}\\
    &=G^{\prime}(X)\left(f^{\prime}(\lambda)-f^{\prime}(1-\lambda)\right)
\end{aligned}
\end{align}
 at  $\lambda=1 / 2$ to give $E_{G}^{\prime}(1 / 2)=0$ .

  \item $E_{G}^{\prime}(\lambda) \neq 0, \forall \lambda \neq 1 / 2$. This property is to ensure that $E_{G}(\lambda)$ has only one extreme point in the range $0 \leq \lambda \leq 1$. From (20), this property requires $G^{\prime}(X) \neq 0$, and $f^{\prime}(\lambda) \neq f^{\prime}(1-$ $\lambda), \forall \lambda \neq 1 / 2$. Hence, $G(X)$ and $f^{\prime}(\lambda)$ have to be strictly increasing or decreasing in their definition domains.

  \item $E_{G}^{\prime \prime}(\lambda)<0$ at $\lambda=1 / 2$. This property ensures that $E_{G}(\lambda)$ is concave downward at the extreme point. From (20), the second derivative of $E_{G}(\lambda)$ reads
  \begin{align}
\label{eqn: 21}
\begin{aligned}
    \frac{\partial^{2} E_{G}(\lambda)}{\partial \lambda^{2}}= & G^{\prime \prime}(X)\left(f^{\prime}(\lambda)-f^{\prime}(1-\lambda)\right)^{2} \\
& +G^{\prime}(X)\left(f^{\prime \prime}(\lambda)+f^{\prime \prime}(1-\lambda)\right).
\end{aligned}
\end{align}
The evaluation of $\partial^{2} E_{G} / \partial \lambda^{2}$ at $\lambda=1 / 2$ gives $E_{G}^{\prime \prime}(1 / 2)=$ $2 G^{\prime}(2 f(1 / 2)) \cdot f^{\prime \prime}(1 / 2)$. Therefore, the downward concavity of $E_{G}(\lambda)$ at $\lambda=1 / 2$ requires
\begin{equation}
\label{eqn: 22}
G^{\prime}(2 f(1 / 2)) \cdot f^{\prime \prime}(1 / 2)<0 .
\end{equation}
Together with condition 4), we then conclude that if $f$ is concave downward, $G$ must be strictly increasing; on the contrary, if $f$ is concave upward, $G$ must be strictly decreasing.
\end{enumerate}
Condition 1) and condition 2) are the basic requirements for the entanglement measure $E_{G}(\lambda)$ to ensure that except for the separable states, $E_{G}(\lambda)$ must be a positive function. Condition 3) to condition 5) require that $E_{G}(\lambda)$ must have a unique extreme point at $\lambda=1 / 2$, which is the global maximum in the range $0 \leq \lambda \leq 1$. When all the five conditions are satisfied, the general entanglement measure $E_{G}(\rho)$ achieves its global maximum at $\lambda=1 / 2$ :
\begin{equation}
\label{eqn: 23}
\max_{\text{all pure} \rho} E_{G}(\rho)=\max _{\lambda \in [0,1]} G(X(\lambda))=G(2 f(1 / 2)) .
\end{equation}
The general entanglement measure $E_{G}(\rho)$ comprises a wide class of entanglement measures, including concurrence, Renyi entropy, and entropy of entanglement, etc.

\noindent \textbf{Example 1:} Concurrence

Concurrence [45] is a common entanglement measure defined by
    \begin{equation}
        \label{eqn: 24a}
        E_{C}(\rho)=\sqrt{2\left(1-\operatorname{Tr}\left(\rho_{M}^{2}\right)\right)}.
    \end{equation}
    The corresponding \(G(X)\) and \(f(\lambda)\) functions for \(E_{C}(\rho)\) are
    \begin{equation*}
        G(X)=\sqrt{2(1-X)}, \quad f(\lambda)=\lambda^{2}
    \end{equation*}
    based on which \(E_{C}(\rho)\) becomes a scalar function of \(\lambda\) :
    \begin{equation}
        \label{eqn: 24b}
        E_{C}(\rho)=G(f(\lambda)+f(1-\lambda))=2 \sqrt{\lambda(1-\lambda)}.
    \end{equation}
It can be checked that $G(X)$ and $f(\lambda)$ satisfy the above five conditions. The maximum of $E_{C}(\rho)$ is found from (23) as $E_{c}\left(\rho_{M}^{*}\right)=G(2 f(1 / 2))=G(1 / 2)=1$, and the downward concavity of $E_{C}(\rho)$ at the extreme point is confirmed by $E_{C}^{\prime \prime}(1 / 2)=2 G^{\prime}(2 f(1 / 2)) \cdot f^{\prime \prime}(1 / 2)=-4<0$.

\noindent \textbf{Example 2:} Renyi entropy

Renyi entropy \cite{horodecki_2009_quantum} is defined by
\begin{equation}
\label{eqn: 25}
E_{\alpha}(\rho)=\frac{1}{1-\alpha} \ln \operatorname{Tr}\left(\rho_{M}^{\alpha}\right), \alpha>0, 
\end{equation}
whose related functions of $G(X)$ and $f(\lambda)$ are
$$
G(X)=\frac{1}{1-\alpha} \ln X, f(\lambda)=\lambda^{\alpha}.
$$
In terms of $\lambda, E_{\alpha}(\rho)$ becomes
$$
E_{\alpha}(\lambda)=\frac{1}{1-\alpha} \ln \left(\lambda^{\alpha}+(1-\lambda)^{\alpha}\right) .
$$
The maximum of $E_{\alpha}(\lambda)$ is $E_{\alpha}(1 / 2)=G(2 f(1 / 2))=\ln 2$ and its downward concavity is confirmed by $E_{\alpha}^{\prime \prime}(1 / 2)=-4 \alpha<0$. A special case of Renyi entropy is the entropy of entanglement $E_{E}(\rho)$, which is the limit value of $E_{\alpha}(\rho)$ at $\alpha=1$ obtained by the L'Hôspital's rule,
\begin{equation}
\label{eqn: 26}
\lim _{\alpha \rightarrow 1} E_{\alpha}(\rho)=-\frac{1}{\ln 2} \operatorname{Tr}\left(\rho_{M} \ln \rho_{M}\right)=E_{E}(\rho) .
\end{equation}

Based on the general entanglement measure $E_{G}(\rho)$, we can now construct a class of Lyapunov functions for entanglement control as
\begin{equation}
\label{eqn：27}
V_{G}(\rho)=\mathcal{N}-E_{G}(\rho)=\mathcal{N}-G\left(\operatorname{Tr}\left(f\left(\rho_{M}\right)\right)\right),
\end{equation}
where $\mathcal{N}=G(2 f(1 / 2))$ is the maximum of $E_{G}(\rho)$ to ensure $V_{G}(\rho) \geq 0$. The Lyapunov function $V_{G}(\rho)$ constructed from $E_{G}(\rho)$ is called the Lyapunov entanglement function (LEF) to highlight its dual role. On the one hand, LEF plays the role of a Lyapunov function and determines the control strategy to make $\dot{V}_{G}(\rho)<0$. On the other hand, it plays the role of an entanglement measure, guiding the control process toward the direction of maximum entanglement. Combining the two roles, the control strategy $\dot{V}_{G}(\rho)<0$ drives $\rho$ to the equilibrium state $\rho_{e q}$ with $V_{G}\left(\rho_{e q}\right)=0$, which then gives $E_{G}\left(\rho_{\text {eq }}\right)=\mathcal{N}$ from \eqref{eqn：27}, indicating that the achieved equilibrium state $\rho_{e q}$ is the MES.

\section{LYAPUNOV CONTROL BASED ON LEF}
\label{sec: Lyapunov_design}
In this section, we derive the control field $u_{k}$ in \eqref{eqn: 6} to make $\dot{V}_{G}(\rho)<0$. First, we discuss the entanglement control of pure states in this section, and then the control of mixed states in Section \ref{sec: Lyapunov_design_mixed}. The first step is to find the time derivative of $V_{G}(\rho)$ from \eqref{eqn：27}:

\begin{align}
\label{eqn: 28}
\begin{aligned}
    \dot{V}_{G}(\rho) & =-G^{\prime}\left(\operatorname{Tr}\left(f\left(\rho_{M}\right)\right)\right) \operatorname{Tr}\left(f^{\prime}\left(\rho_{M}\right) \dot{\rho}_{M}\right) \\
& =i G^{\prime}\left(\operatorname{Tr}\left(f\left(\rho_{M}\right)\right)\right) \operatorname{Tr}\left(f^{\prime}\left(\rho_{M}\right) \cdot(H \rho-\rho H)_{M}\right)    
\end{aligned}
\end{align}
where $\dot{\rho}$ is given by \eqref{eqn: 6} and $(\cdot)_{M}$ denotes the partial trace operation. Next, we use the expression of Hamiltonian $H$ under the interaction picture to rewrite \eqref{eqn: 28} as:
\begin{align}
\label{eqn: 29}
\begin{aligned}
    \dot{V}_{G}(\rho)= & i G^{\prime}\left(\operatorname{Tr}\left(f\left(\rho_{M}\right)\right)\right) \\
& \cdot \sum_{k} u_{k} \operatorname{Tr}\left(f^{\prime}\left(\rho_{M}\right) \cdot\left(A_{k} \rho-\rho A_{k}\right)_{M}\right).
\end{aligned}
\end{align}
On designing the control law $u_{k}$ to render $\dot{V}_{G}(\rho)<0$, the following theorem is helpful.

\begin{proposition}
\label{prop: 1}
    $\operatorname{Tr}\left(f^{\prime}\left(\rho_{M}\right) \cdot\left(A_{k} \rho-\rho A_{k}\right)_{M}\right)$ is an imaginary number.
\end{proposition}
\begin{proof}
Since $\rho_{M}$ is Hermitian, $f^{\prime}\left(\rho_{M}\right)$ is also Hermitian and can be expressed generally as
\begin{align}
    \label{eqn: 30}
    f^{\prime}\left(\rho_{M}\right)=\left[\begin{array}{ll}
q_{11} & q_{12}  \\
q_{12}^{*} & q_{22}
\end{array}\right],
\end{align}
where $q_{11}$ and $q_{22}$ are real numbers, and $q_{12}$ is a complex number. With the Hermitian property of $A_{k}$ and $\rho$, we have
\begin{equation}
\label{eqn: 31}
\left(A_{k} \rho-\rho A_{k}\right)^{\dagger}=\rho A_{k}-A_{k} \rho=-\left(A_{k} \rho-\rho A_{k}\right) .
\end{equation}
In other words, $A_{k} \rho-\rho A_{k}$ is a skew-Hermitian matrix. In the next step of the proof, we apply the rules of trace operation:
$\operatorname{Tr}(A+B)=\operatorname{Tr}(A)+\operatorname{Tr}(B)$ and $\operatorname{Tr}(A B)=\operatorname{Tr}(B A)$ to obtain $\operatorname{Tr}\left(A_{k} \rho-\rho A_{k}\right)=\operatorname{Tr}\left(A_{k} \rho\right)-\operatorname{Tr}\left(\rho A_{k}\right)=0$. Combining the skew-Hermitian and zero-trace properties of $A_{k} \rho-\rho A_{k}$, we now can express $\left(A_{k} \rho-\rho A_{k}\right)_{M}$ explicitly as
\begin{align}
\label{eqn: 32}
    \left(A_{k} \rho-\rho A_{k}\right)_{M}=\left[\begin{array}{ll}
\eta_{k} & \zeta_{k}  \\
-\zeta_{k}^{*} & -\eta_{k}
\end{array}\right], k=1,2, \cdots, m,
\end{align}
where $\eta_{k}$ is a pure imaginary number, and $\zeta_{k}$ is a complex number. Therefore, the combination of \eqref{eqn: 30} and \eqref{eqn: 32} yields
\begin{align}
\label{eqn: 33}
\begin{aligned}
&\operatorname{Tr}\left(f^{\prime}\left(\rho_{M}\right) \cdot\left(A_{k} \rho-\rho A_{k}\right)_{M}\right)
\\
&=\operatorname{Tr}\left(\left[\begin{array}{ll}q_{11} & q_{12} \\ q_{12}^{*} & q_{22}\end{array}\right]\left[\begin{array}{cc}\eta_{k} & \zeta_{k} \\ -\zeta_{k}^{*} & -\eta_{k}\end{array}\right]\right)\\
&=\left(q_{11}-q_{22}\right) \eta_{k}+q_{12}^{*} \zeta_{k}-\left(q_{12}^{*} \zeta_{k}\right)^{*}.   
\end{aligned}
\end{align}
Noting that $q_{11}-q_{22}$ is real, and $\eta_{k}, q_{12}^{*} \zeta_{k}-\left(q_{12}^{*} \zeta_{k}\right)^{*}$ are pure imaginary, we then prove $\operatorname{Tr}\left(f^{\prime}\left(\rho_{M}\right) \cdot\left(A_{k} \rho-\rho A_{k}\right)_{M}\right)$ to be a pure imaginary number.
\end{proof}

The other factor affecting the sign of $\dot{V}_{G}(\rho)$ in \eqref{eqn: 29} is $G^{\prime}\left(\operatorname{Tr}\left(f\left(\rho_{M}\right)\right)\right)$. We have shown in Section III that for a qualified entanglement measure $E_{G}(\rho)=G\left(\operatorname{Tr}\left(f\left(\rho_{M}\right)\right)\right)$, the function $G(X)$ must be either strictly increasing, or strictly decreasing. In either case, it can be sure that $G^{\prime}(X)$ will not change sign in its domain of definition.

With Proposition \ref{prop: 1} and the monotonic property of $G^{\prime}(X)$, we now can design the Lyapunov control law $u_{k}$ in terms of a new variable
\begin{equation}
\label{eqn: 34}
x_{k}=i \cdot \operatorname{Tr}\left(f^{\prime}\left(\rho_{M}\right) \cdot\left(A_{k} \rho-\rho A_{k}\right)_{M}\right) .
\end{equation}
According to Proposition \ref{prop: 1}, $x_{k}$ is a real variable and can be physically realized. Let $h_{k}\left(x_{k}\right)$ be a function of $x_{k}$ satisfying the relation
\begin{equation}
\label{eqn: 35}
h_{k}\left(x_{k}\right) \cdot x_{k} \geq 0, h_{k}\left(x_{k}\right)=0 \text {, iff } x_{k}=0 .
\end{equation}
It is clear that the curve $y_{k}=h_{k}\left(x_{k}\right)$ passes through the origin of the $x_{k}-y_{k}$ plane and is located in the first or third quadrant. Then, the real function $h_{k}\left(x_{k}\right)$ serves as a feedback signal in the proposed control field
\begin{equation}
\label{eqn: 36}
u_{k}=-\operatorname{sgn}\left(G^{\prime}(X)\right) r_{k} h_{k}\left(x_{k}\right), 
\end{equation}
where $r_{k}$ is a positive gain to adjust the control amplitude. Applying the control law \eqref{eqn: 36} to \eqref{eqn: 29}, we achieve the goal of Lyapunov control
\begin{equation}
\label{eqn: 37}
\dot{V}_{G}(\rho)=-\left|G^{\prime}(X)\right| \sum_{k} r_{k} h_{k}\left(x_{k}\right) \cdot x_{k} \leq 0 ,
\end{equation}
by noting $h_{k}\left(x_{k}\right) \cdot x_{k} \geq 0$ from \eqref{eqn: 35}. Our next task is to show that $\dot{V}_{G}(\rho)=0$ occurs only at the equilibrium state $\rho=\rho_{e q}$ and $\dot{V}_{G}(\rho)<0, \forall \rho \neq \rho_{e q}$. The dual role of the $\operatorname{LEF} V_{G}(\rho)$ ensures that the minimum of $V_{G}$ and the maximum of $E_{G}$ are achieved simultaneously at $\rho=\rho_{e q}$.
\begin{theorem}
\label{thm: 1}
    Under the Lyapunov control law \eqref{eqn: 36}, a pure bipartite state $\rho(t)$ following the time evolution \eqref{eqn: 6} asymptotically converges to the equilibrium state $\rho_{e q}$ such that $V_{G}(\rho)$ reaches its minimum $V_{G}\left(\rho_{e q}\right)=0$ and the general entanglement measure $E_{G}(\rho)$ reaches its maximum $E_{G}\left(\rho_{e q}\right)=$ $\mathcal{N}$.
\end{theorem}
\begin{proof}
  According to Barbalat's lemma \cite{slotine1991applied}, if $\dot{V}_{G}(\rho)$ is is uniformly continuous, the condition $\dot{V}_{G}(\rho) \leq 0$ guarantees that the state trajectory $\rho(t)$ converges to the invariant set
\begin{equation}
\label{eqn: 38}
\Omega_{G}=\left\{\rho \mid \dot{V}_{G}(\rho)=0, \rho \in \mathcal{H}_{A B}\right\}.
\end{equation}
Firstly, \( \dot{V}_{G}(\rho) \) is uniformly continuous since the functions \( G \) and \( f \) are twice continuously differentiable, and the differential term for the reduced density matrix \( \rho_M \) can be rewritten as \( (H \rho - \rho H)_M \), which is naturally finite. Thus, \( \ddot{V}_{G}(\rho) \) is bounded, implying that \( \dot{V}_{G}(\rho) \) is uniformly continuous. Then, according to the property of $h_{k}\left(x_{k}\right) \cdot x_{k} \geq 0$ in \eqref{eqn: 35}, the condition of $\dot{V}_{G}(\rho)=-\left|G^{\prime}(X)\right| \sum_{k} r_{k} h_{k}\left(x_{k}\right) \cdot x_{k}=0$ occures only at $x_{k}=$ $0, k=1,2, \cdots, m$, where $m$ is the number of control field $u_{k}$ used in \eqref{eqn: 6}. With $x_{k}=0$, \eqref{eqn: 36} gives $u_{k}=-\operatorname{sgn}\left(G^{\prime}(X)\right)$. $r_{k} h_{k}\left(x_{k}\right)=0$, because of $h_{k}(0)=0$. Applying $u_{k}=0$ to \eqref{eqn: 6} then yields the equilibrium condition $\dot{\rho}(t)=0$. Therefore, the invariant set $\Omega_{G}$ defined in \eqref{eqn: 38} contains the equilibrium states $\rho_{e q}$ of the von Neumann equation \eqref{eqn: 6}, and the state trajectory $\rho(t)$ converges asymptotically to $\rho_{e q}$ such that $\dot{V}_{G}\left(\rho_{e q}\right)=0$.

The proof of the rest of the theorem is about the properties of the equilibrium state $\rho_{e q}$, which can be derived from the equilibrium condition $x_{k}=0$. In terms of \eqref{eqn: 33}, the equilibrium condition $x_{k} \triangleq i \operatorname{Tr}\left(f^{\prime}\left(\rho_{M}\right) \cdot\left(A_{k} \rho-\rho A_{k}\right)_{M}\right)=0$ can be expressed by
\begin{align}
\label{eqn: 39}
    \left(q_{11}-q_{22}\right) \eta_{k}+q_{12}^{*} \zeta_{k}-\left(q_{12}^{*} \zeta_{k}\right)^{*}=0, k=1,2, \cdots, m .
\end{align}
Eq. \eqref{eqn: 39} has to be satisfied for all $\eta_{k}$ and $\zeta_{k}$, in order to achieve the condition $x_{k}=0$, and the only possibility is $q_{11}=q_{22}$ and $q_{12}=0$, which in turn is substituted into \eqref{eqn: 30} to yield
\begin{align}
\label{eqn: 40}
    f^{\prime}\left(\left(\rho_{e q}\right)_{M}\right)=\left[\begin{array}{ll}
q_{11} & q_{12} \\
q_{12}^{*} & q_{22}
\end{array}\right]=\left[\begin{array}{ll}
q_{11} & 0 \\
0 & q_{11}
\end{array}\right] .
\end{align}
Because $f^{\prime}$ is either strictly increasing or strictly decreasing as proved in Section \ref{sec: Lyapunov entangled function}, its inverse function $\left[f^{\prime}\right]^{-1}$ always exists and the equilibrium state can be solved as
\begin{align}
    \label{eqn: 41}
\begin{aligned}
       \left(\rho_{e q}\right)_{M}&=\left[\begin{array}{ll}
{\left[f^{\prime}\right]^{-1}\left(q_{11}\right)} & 0  \\
0 & {\left[f^{\prime}\right]^{-1}\left(q_{11}\right)}
\end{array}\right]\\
&=\left[\begin{array}{ll}
1 / 2 & 0 \\
0 & 1 / 2
\end{array}\right],
\end{aligned}
\end{align}
where the identity $\operatorname{Tr}\left(\rho_{M}\right)=1$ has been used to determine the value of $\left[f^{\prime}\right]^{-1}\left(q_{11}\right)$. Comparing \eqref{eqn: 41} with \eqref{eqn: 19}, we obtain the main result of this theorem that the equilibrium state $\rho_{e q}$ achieved by the Lyapunov control law \eqref{eqn: 36} is identical to $\rho_{\text{MES}}$. Because of $E_{G}\left(\rho_{\text{MES}}\right)=\mathcal{N}$, the Lyapunov function evaluated at the equilibrium state becomes $V_{G}\left(\rho_{e q}\right)=\mathcal{N}-E_{G}\left(\rho_{e q}\right)=$ $\mathcal{N}-E_{G}\left(\rho_{\text{MES}}\right)=0$.  
\end{proof}

Theorem \ref{thm: 1} shows that the proposed Lyapunov control law \eqref{eqn: 36} can drive the quantum state to the MES, which maximizes the general entanglement measure $E_{G}(\rho)$. It is noted that the MESs generated by the Lyapunov entanglement control are not limited to Bell states but contain all the $\rho_{\text{MES}}$ that achieve $E_{G}\left(\rho_{\text{MES}}\right)=\mathcal{N}$.

\section{Numerical verification of maximum entanglement control}
\label{sec: numerical verification of maximum entanglement control}
In this section, we numerically verify the Lyapunov entanglement control method derived in the previous section. We consider a model representing two atoms each located in a remote cavity connected by a closed-loop optical fiber. One of the two atoms is given a coherent input field of amplitude $A_{m}$, and the output of each cavity enters the other. By eliminating the radiation field, the internal Hamiltonian is chosen as $H_{0}=2 J \sigma_{z} \otimes \sigma_{z}$, where the spin-spin coupling constant $J$ changes with the frequency of the applied radiation field and $J=0.5$ is used in the computation.

The control Hamiltonian $H_{c}=\sum_{k=1}^{3} H_{k} u_{k}(t)$ is synthesized by a local laser and the coupling Hamiltonian $H_{k}$ is a combination of Pauli matrices $\sigma_{x}, \sigma_{y}, \sigma_{z}$. Here we choose $H_{1}=\sigma_{x} \otimes \sigma_{y}+\sigma_{z} \otimes \sigma_{z}, H_{2}=\sigma_{x} \otimes \sigma_{z}+\sigma_{z} \otimes \sigma_{x}$, and $H_{3}=\sigma_{y} \otimes \sigma_{z}+\sigma_{z} \otimes \sigma_{y}$. With the given $H_{k}$, the time evolution of the density matrix is described by \eqref{eqn: 6}, and the control field $u_{k}$ is given by \eqref{eqn: 36}, where the feedback signal is chosen to be the simplest form $h_{k}\left(x_{k}\right)=x_{k}$ with gain $r_{k}=5$. The density matrix $\rho$ for pure states is described by $\rho=|\psi\rangle\langle\psi|$ with the quantum state $|\psi(t)\rangle$ expressed in terms of the basis as
\begin{equation}
\label{eqn: 42}
|\psi(t)\rangle=\alpha|00\rangle+\beta|01\rangle+\gamma|10\rangle+\delta|11\rangle.
\end{equation}
The reduced maxtix $\rho_{M}$ of the pure state $\rho=|\psi\rangle\langle\psi|$ can be computed in terms of the coefficients of $|\psi(t)\rangle$ as
\begin{align}
\label{eqn: 43}
    \rho_{M}=\left[\begin{array}{ll}
|\alpha|^{2}+|\beta|^{2} & \beta \bar{\delta}+\alpha \bar{\gamma} \\
\delta \bar{\beta}+\gamma \bar{\alpha} & |\gamma|^{2}+|\delta|^{2}
\end{array}\right],
\end{align}
which is then used in $E_{G}(\rho)=G\left(\operatorname{Tr}\left(f\left(\rho_{M}\right)\right)\right)$ to compute the entanglement measure.

Once the control process is activated, it will asymptotically converge to an MES regardless of the initial states. What we are interested in is, from what initial states, the obtained MES just has the form of Bell states, i.e.,
\begin{align}
\label{eqn: 44}
\begin{aligned}
     \left|\beta_{00}\right\rangle &= \frac{1}{\sqrt{2}}(\left|00\right\rangle + \left|11\right\rangle), & \left|\beta_{01}\right\rangle &= \frac{1}{\sqrt{2}}(\left|00\right\rangle - \left|11\right\rangle), \\
    \left|\beta_{10}\right\rangle &= \frac{1}{\sqrt{2}}(\left|01\right\rangle + \left|10\right\rangle), & \left|\beta_{11}\right\rangle &= \frac{1}{\sqrt{2}}(\left|01\right\rangle - \left|10\right\rangle) .
\end{aligned}
\end{align}
Firstly, we consider the initial state $\left|\psi_{0}\right\rangle=[(1+$ $\epsilon)\left|\beta_{00}\right\rangle+\left|\beta_{01}\right\rangle ]/ \sqrt{2+2\epsilon+\epsilon^2}\approx|00\rangle$, which has a small perturbation $\epsilon$ from the separable state $|00\rangle$ to examine the influence of the perturbation of initial states on the convergence to Bell states.

\begin{figure}[htbp]
\centering
\includegraphics[width=\columnwidth]{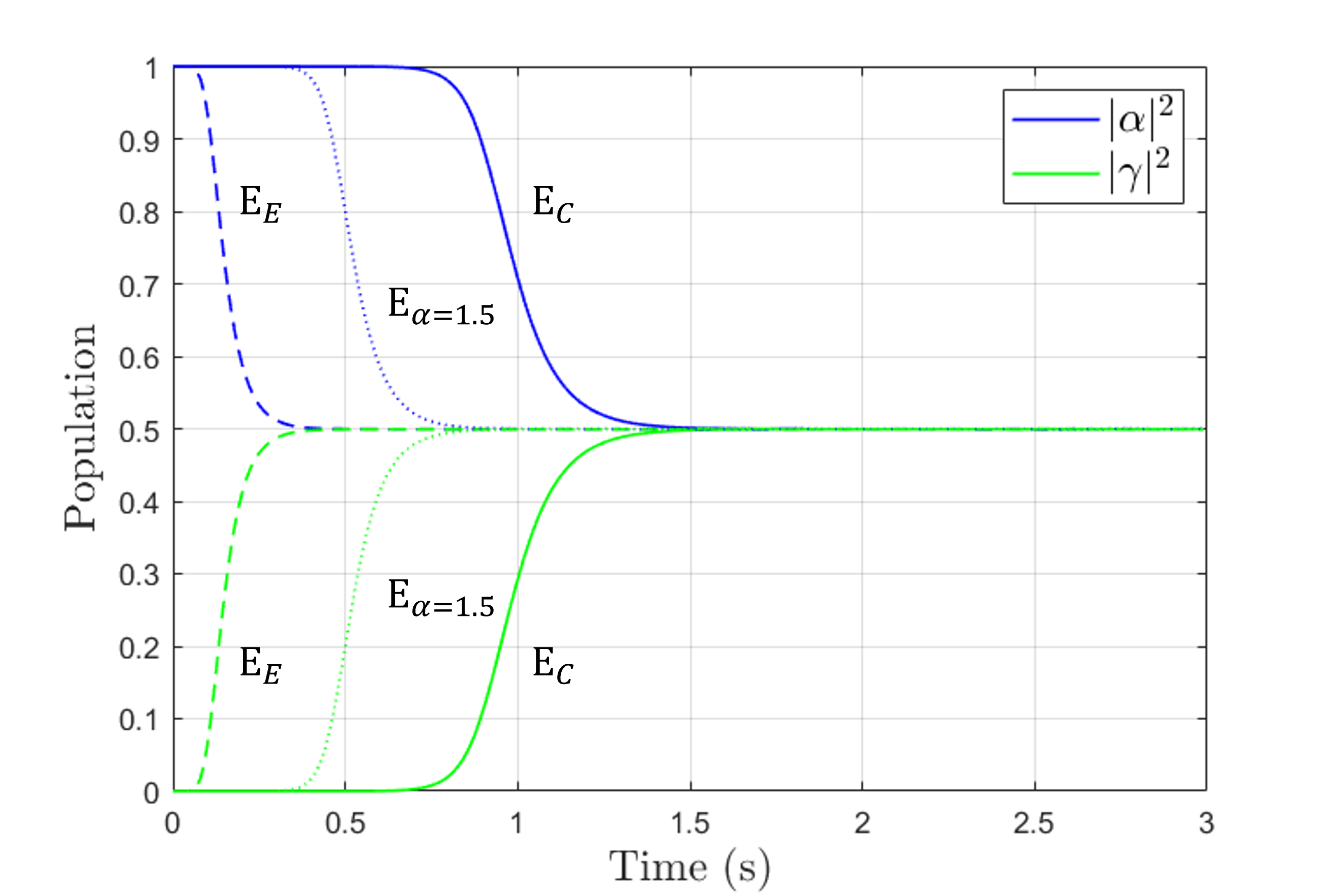}
\caption{Asymptotic convergence of the state population from the initial state $|00\rangle$ to the Bell state $\left|\beta_{00}\right\rangle$ based on three entanglement measures: concurrence $E_{C}(\rho)$, entropy of entanglement $E_{E}(\rho)$, and Renyi entropy $E_{\alpha}(\rho)$ with $\alpha=1.5$. In this context, $\alpha$ and $\gamma$ are defined in \eqref{eqn: 42}, while $\beta$ and $\delta$ are always $0$.}
\label{fig: 1}
\end{figure}
Fig. \ref{fig: 1} shows the time response of the population in each basis state under the maximum entanglement control by using three entanglement measures: concurrence $E_{C}(\rho)$, entropy of entanglement $E_{E}(\rho)$, and Renyi entropy $E_{\alpha}(\rho)$ with $\alpha=1.5$, as introduced in the previous section. It can be seen that the quantum states starting from $|00\rangle$ all converge asymptotically to the Bell state $\left|\beta_{00}\right\rangle$ by using three different entanglement measures. The different convergent speeds observed from Fig. \ref{fig: 1} can be explained by the time response of the related control fields shown in Fig. 2. The control field generated by the entropy of entanglement $E_{E}(\rho)$ activates first and drives the quantum state to the Bell state faster than that by using the control field generated by the concurrence $E_{C}(\rho)$, which is the last of the three control fields to be activated. Although the three control fields are activated at different moments, their magnitudes are the same.

The initial state $\left|\psi_{0}\right\rangle=[(1+\epsilon)\left|\beta_{00}\right\rangle+\left|\beta_{01}\right\rangle]/ \sqrt{2+2\epsilon+\epsilon^2} \approx|00\rangle$ considered previously has a slightly larger weight on $\left|\beta_{00}\right\rangle$ than $\left|\beta_{01}\right\rangle$ and causes the state to converge to $\left|\beta_{00}\right\rangle$. 
Now we add the perturbation $\epsilon$ to $\left|\beta_{01}\right\rangle$, instead of $\left|\beta_{00}\right\rangle$, to form a different initial state $\left|\psi_{0}\right\rangle=[\left|\beta_{00}\right\rangle+(1+\epsilon)\left|\beta_{01}\right\rangle ]/ \sqrt{2+2\epsilon+\epsilon^2} \approx |00\rangle$. By applying the same Lyapunov entanglement control, the terminal state turns out to be $\left|\beta_{01}\right\rangle$. The terminal Bell state is highly sensitive to the quantum state's departure from the initial state. Table 1 lists the initial states with different perturbation and their corresponding terminal states under the same Lyapunov entanglement control. The results of this table show that if the initial state is close to the four basis states, the achieved MES appears to be one of the Bell states. Furthermore, perturbing the initial state towards a specific Bell state directs the quantum state evolution towards that Bell state.

\begin{figure}[htbp]
\includegraphics[width=\columnwidth]{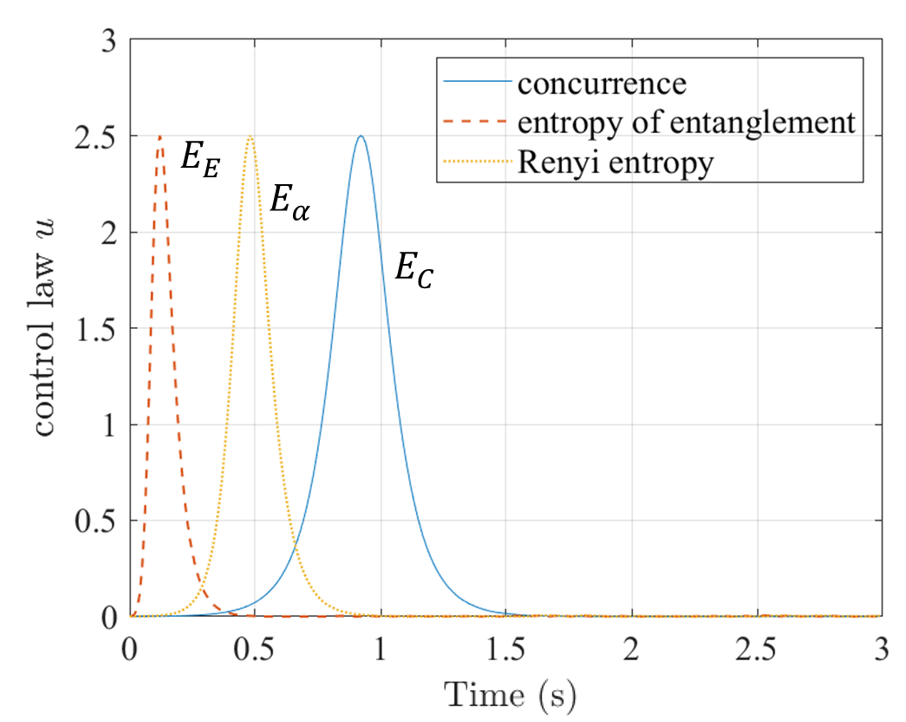}
\caption{The time responses of the control field used to drive the state from $|00\rangle$ to $\left|\beta_{00}\right\rangle=(|00\rangle+|11\rangle) / \sqrt{2}$ based on three entanglement measures.}
\label{fig: 2}
\end{figure}

\begin{table}[htbp]
\caption{Initial states and the related final states under the same Lyapunov entanglement control. All initial deviations are normalized by \(\sqrt{2+2\epsilon+\epsilon^2}\).}
    \label{tab: 1}
    \begin{center}
    \begin{tabular}{|c|c|c|c|}
    \hline
    \multicolumn{2}{|c|}{Initial state} & Initial deviation & Final state \\
    \hline
    1 & $\approx |00\rangle$ & $(1+\epsilon)\left|\beta_{00}\right\rangle+\left|\beta_{01}\right\rangle$ & $\left|\beta_{00}\right\rangle$ \\
    \hline
    2 & $\approx |00\rangle$ & $\left|\beta_{00}\right\rangle+(1+\epsilon)\left|\beta_{01}\right\rangle$ & $\left|\beta_{01}\right\rangle$ \\
    \hline
    3 & $\approx |11\rangle$ & $(1+\epsilon)\left|\beta_{00}\right\rangle-\left|\beta_{01}\right\rangle$ & $\left|\beta_{00}\right\rangle$ \\
    \hline
    4 & $\approx |11\rangle$ & $\left|\beta_{00}\right\rangle-(1+\epsilon)\left|\beta_{01}\right\rangle$ & $\left|\beta_{01}\right\rangle$ \\
    \hline
    5 & $\approx |01\rangle$ & $(1+\epsilon)\left|\beta_{10}\right\rangle+\left|\beta_{11}\right\rangle$ & $\left|\beta_{10}\right\rangle$ \\
    \hline
    6 & $\approx |01\rangle$ & $\left|\beta_{10}\right\rangle+(1+\epsilon)\left|\beta_{11}\right\rangle$ & $\left|\beta_{11}\right\rangle$ \\
    \hline
    7 & $\approx |10\rangle$ & $(1+\epsilon)\left|\beta_{10}\right\rangle-\left|\beta_{11}\right\rangle$ & $\left|\beta_{10}\right\rangle$ \\
    \hline
    8 & $\approx |10\rangle$ & $\left|\beta_{10}\right\rangle-(1+\epsilon)\left|\beta_{11}\right\rangle$ & $\left|\beta_{11}\right\rangle$ \\
    \hline
    \end{tabular}
    \end{center}
\end{table}

\begin{figure}[htbp]
\centering
\includegraphics[width=\columnwidth]{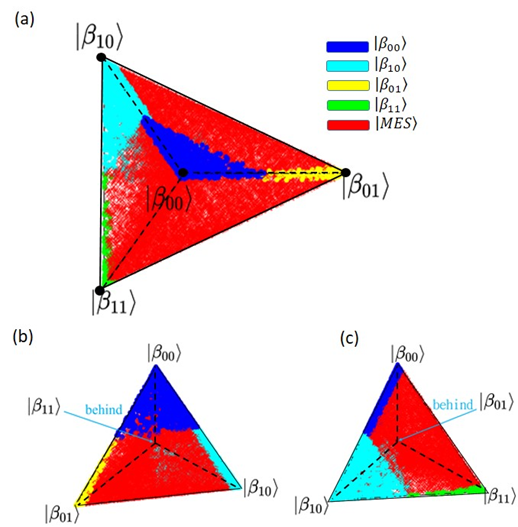}
\caption{Distribution of initial states converging to different Bell states on a tetrahedron. Blue, cyan, yellow, and green regions represent initial states converging to $\left.\left|\beta_{00}\right\rangle, \beta_{10}\right\rangle,\left|\beta_{01}\right\rangle$, and $\left|\beta_{11}\right\rangle$, respectively. The red region covers those initial states converging to the Bell equivalent states. (b) and (c) are the results of observing the tetrahedron in (a) from different orientations.}
\label{fig: 3}
\end{figure}

To understand the global convergence range of Bell states, we select a large number of initial states at random and identify their terminal states under the same maximum entanglement control. For this purpose, a quantum state is represented as a linear combination of four Bell states
\begin{equation}
\label{eqn: 45}
|\psi(t)\rangle=\beta_{\alpha}\left|\beta_{00}\right\rangle+\beta_{\beta}\left|\beta_{01}\right\rangle+\beta_{\gamma}\left|\beta_{10}\right\rangle+\beta_{\delta}\left|\beta_{11}\right\rangle ,
\end{equation}
where the coefficients satisfy the normalization condition. Graphically, the four Bell states can be thought of as the four vertices of a regular tetrahedron, and the coefficient set $\left\{\beta_{\alpha}, \beta_{\beta}, \beta_{\gamma}, \beta_{\delta}\right\}$ determines the position of the corresponding quantum state in the tetrahedron, as shown in Fig. \ref{fig: 3}.

We make use of different colors to distinguish regions converging to different Bell states in such a way that a blue dot represents an initial state, which converges to the Bell state $\left|\beta_{00}\right\rangle$, and cyan, yellow, and green dots represent those initial states converging to $\left|\beta_{10}\right\rangle,\left|\beta_{01}\right\rangle$, and $\left|\beta_{11}\right\rangle$, respectively. The red dots, which cover most of the tetrahedron, correspond to the initial states converging to the MESs not in the form of Bell states, which are called Bell equivalent states, i.e., they are equivalent to Bell states under local unitary transformation.

It can be seen from Fig. \ref{fig: 3} that the initial states close to the Bell states at the four corners of the tetrahedron tend to converge to their nearby Bell states. Besides the regions close to the four corners, initial states distributing along the line connecting $\left|\beta_{00}\right\rangle$ and $\left|\beta_{01}\right\rangle$ and the line connecting $\left|\beta_{10}\right\rangle$ and $\left|\beta_{11}\right\rangle$ also tend to converge to the Bell states. This finding is consistent with the result of Table \ref{tab: 1}, where the combination of $\left|\beta_{00}\right\rangle$ and $\left|\beta_{01}\right\rangle$ yields the first four Bell states and the combination of $\left|\beta_{10}\right\rangle$ and $\left|\beta_{11}\right\rangle$ yields the next four Bell states in Table \ref{tab: 1}.

The Lyapunov entanglement control ensures that all the initial states converge to the MESs, which include Bell states and Bell equivalent states, as shown in Fig. \ref{fig: 3}. The MES for a bipartite pure state has a general expression as
\begin{equation}
\label{eqn: 46}
\left|M E S_{ \pm}\right\rangle=\frac{1}{\sqrt{2}}\left(p|00\rangle+q|01\rangle \mp q^{*}|10\rangle \pm p^{*}|11\rangle\right) . 
\end{equation}
This general expression can be confirmed by substituting the coefficients of $\left|M E S_{ \pm}\right\rangle$ into the reduced matrix \eqref{eqn: 43} to yield
\begin{align}
\label{eqn: 47}
\begin{aligned}
     \rho_{M}&=\left[\begin{array}{ll}
|\alpha|^{2}+|\beta|^{2} & \beta \bar{\delta}+\alpha \bar{\gamma} \\
\delta \bar{\beta}+\gamma \bar{\alpha} & |\gamma|^{2}+|\delta|^{2}
\end{array}\right]\\
&=\left[\begin{array}{ll}
1 / 2 & 0 \\
0 & 1 / 2
\end{array}\right] ,
\end{aligned}
\end{align}
which recovers the reduced matrix of the MES given by \eqref{eqn: 19}. If either $p$ or $q$ is equal to zero in \eqref{eqn: 46}, $\left|M E S_{+}\right\rangle$ becomes Bell state; if both $p$ and $q$ are not equal to zero, $\left|M E S_{ \pm}\right\rangle$ is a Bell equivalent state.

\begin{figure}[htbp!]
\centering
    \includegraphics[width=0.96\linewidth]{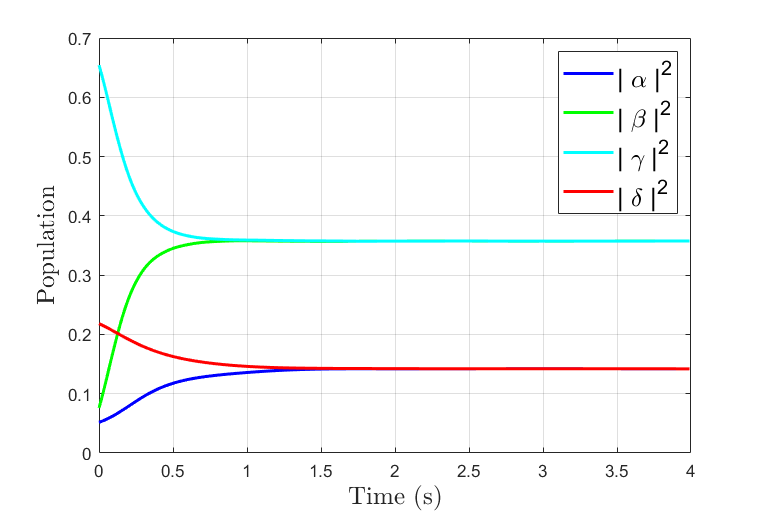}
    \text{(a) Concurrence (solid line)}
    \includegraphics[width=0.96\linewidth]{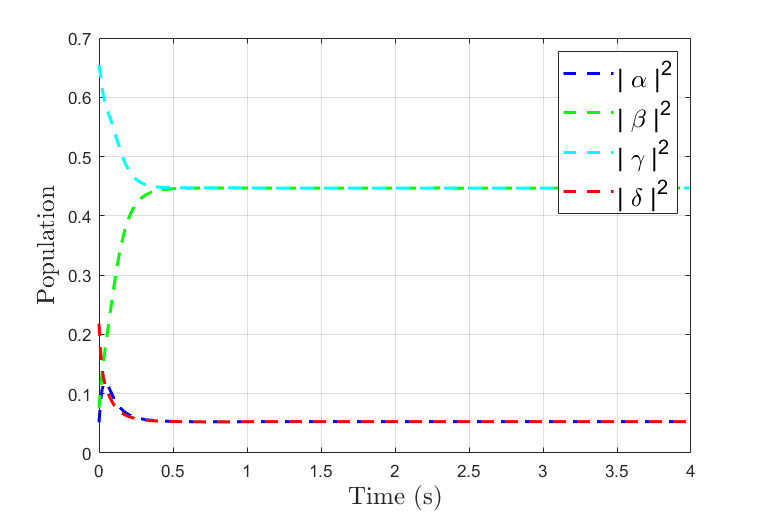}
    \text{(b) Entropy of entanglement (dashed line)}
    \includegraphics[width=0.96\linewidth]{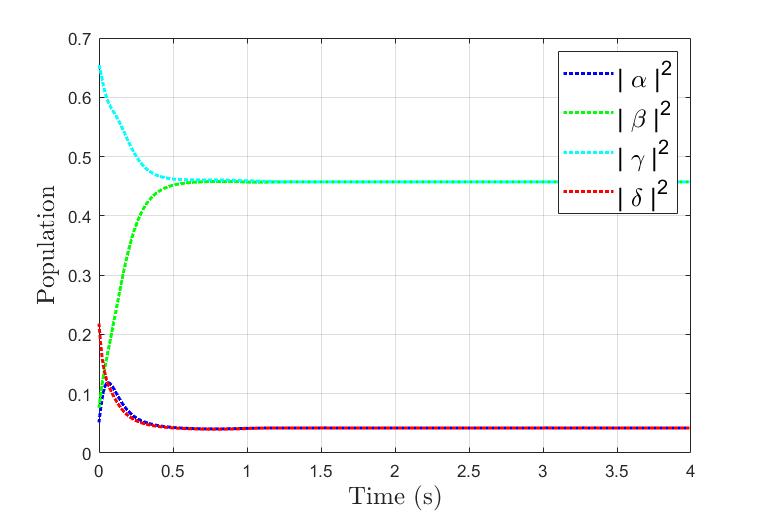}
    \text{(c) Renyi entropy with $\alpha=1.5$ (dotted line)}
\caption{Time evolution of the component populations from an initial state to a terminal Bell equivalent state with components $|\alpha|^{2}=|\delta|^{2}$ and $|\beta|^{2}=|\gamma|^{2}$ for three entanglement measures: concurrence, entropy of entanglement, and Renyi entropy with $\alpha=1.5$.}
\label{fig: 4}
\end{figure}

When the Lyapunov entanglement control converges to a Bell equivalent state, all the populations of the four basis states are not zero according to \eqref{eqn: 46}. This is different from the case of a Bell state, which has only two basis states with non-zero populations. The convergence of Lyapunov entanglement control to a Bell equivalent state is shown in Fig. \ref{fig: 4}, where the initial state is randomly chosen from the red region in Fig. \ref{fig: 3} so that the achieved terminal state is a Bell equivalent state.

\section{Lyapunov entanglement control for mixed states}
\label{sec: Lyapunov_design_mixed}
The existing approach to generating MES as a problem of state transfer becomes invalid for bipartite mixed states because an analytical expression for the maximally entangled mixed states (MEMS) is still unknown. However, the present method can be used to automatically search for the MEMS without specifying it in advance. For a bipartite mixed state $\rho$, there are many different ways to decompose it into the component pure states as $\rho=\sum_{k} p_{k}\left|\psi_{k}\right\rangle\left\langle\psi_{k}\right|$. The degree of entanglement of a mixed state $\rho$ then can be measured by the concurrence of its component pure states as
\begin{equation}
\label{eqn: 48}
E_{G}(\rho)=\min _{\left\{p_{k},\left|\psi_{k}\right\rangle\right\}} \sum_{k} p_{k} E_{G}\left(\left|\psi_{k}\right\rangle\left\langle\psi_{k}\right|\right),
\end{equation}
where the minimization is over all possible ways of decomposition of $\rho$. Any qualified mixed-state entanglement measure can be used in the maximum entanglement control, and the MEMS obtained by them is the same. The reason is that there exists a monotonic mapping between any two qualified entanglement measures. For example, the monotonic mapping between the concurrence $E_{C}$ and the general entanglement measure $E_{G}$ can be derived from \eqref{eqn: 24b} as
\begin{align}
\label{eqn: 49}
    \begin{aligned}
        E_{G}&=G(f(\lambda)+f(1-\lambda)), \\
        \lambda&=\left(1+\sqrt{1-E_{c}^{2}}\right) / 2, 
    \end{aligned}
\end{align}
where $G$ is a monotonic function as verified in Section III. Therefore, the result of maximum entanglement control based on the measure $E_{G}$ is identical to that based on the measure $E_{C}$. We employ the concurrence $E_{C}$ as a demonstration of applying Lyapunov entanglement control to two-qubit mixed states.

The MEMS over the entire $\mathcal{H}_{A B}$ space is still unknown in the literature. Ishizaka and Hiroshima \cite{ishizaka_2000_maximally} proposed a special class of MEMS for two-qubit systems, whose concurrence $E_{C}$ is maximized over all mixed states with given spectrum $\left\{\lambda_{1}, \lambda_{2}, \lambda_{3}, \lambda_{4}\right\}$. This class of MEMS can be generated by applying unitary local transformations to the whole MEMS
\begin{align}
\label{eqn: 50}
\begin{aligned}
    \rho_{\text{MEMS}} & =\lambda_{1}\left|\beta_{11}\right\rangle\left\langle\beta_{11}\left|+\lambda_{2}\right| 00\right\rangle\langle 00| \\
& +\lambda_{3}\left|\beta_{10}\right\rangle\left\langle\beta_{10}\left|+\lambda_{4}\right| 11\right\rangle\langle 11|, 
\end{aligned}
\end{align}
where $\lambda_{i}$ are the eigenvalues of $\rho_{\text {MEMS}}$ in decreasing order with $\lambda_{1}+\lambda_{2}+\lambda_{3}+\lambda_{4}=1$. All the MEMSs in this class have the same concurrence given by
\begin{align}
\label{eqn: 51}
    \begin{aligned}
        &E_{c}\left(\rho_{\text{MEMS}}\right) =E_{c}^{*} \\
        &=\max \left\{0, \lambda_{1}-\lambda_{3}-2 \sqrt{\lambda_{2} \lambda_{4}}\right\},
    \end{aligned}
\end{align}
which is proved to be the maximum concurrence that can be achieved for all mixed states with given spectrum $\left\{\lambda_{1}, \lambda_{2}, \lambda_{3}, \lambda_{4}\right\}$. The role of $\rho_{\text{MEMS}}$ in the mixed state is similar to that of the Bell state in the pure state; however, no quantum control has been proposed to generate this Bell-like mixed state till now. The Lyapunov entanglement control developed in Section IV is particularly suitable for this task because the operation involved in von Neumann equation \eqref{eqn: 4} is just a unitary transformation for $\rho$ so that its spectrum remains unchanged during the control process.

Like \eqref{eqn：27}, the LEF for a mixed state is chosen as
\begin{align}
\label{eqn: 52}
    \begin{aligned}
        V_{c}(\rho)&=\mathcal{N}-E_{c}(\rho)\\
        &=\mathcal{N}-\min _{\left\{p_{k}, \psi_{k}\right\}} \sum_{k} p_{k} E_{c}\left(\left|\psi_{k}\right\rangle\left\langle\psi_{k}\right|\right),
    \end{aligned}
\end{align}
where $\mathcal{N}$ is a constant, which can be set to the maximum of $E_{c}(\rho)$, i.e., $\mathcal{N}=E_{c}^{*}$, to ensure $V_{c}(\rho) \geq 0$. However, the MEMS is determined by the condition $\dot{V}_{c}(\rho)=0$, which is independent of the actual value of $\mathcal{N}$. In other words, the maximum entanglement measure $E_{c}^{*}$ need not be specified in advance in the Lyapunov entanglement control. Once $\rho_{\text{MEMS}}$ is obtained by the entanglement control, $E_{c}\left(\rho_{\text{MEMS}}\right)$ automatically gives the value of $E_{c}^{*}$. The analytical expression of $E_{c}^{*}$ introduced in \eqref{eqn: 51} is only to compare with the $E_{c}^{*}$ obtained by the proposed entanglement control.

For a given mixed state $\rho$, the evaluation of concurrence $E_{c}(\rho)$ involves a minimum decomposition process \eqref{eqn: 48}, which causes difficulty in expressing $\dot{V}_{c}(\rho)$ as an explicit function of $\rho$. Fortunately, this difficulty can be overcome by the method of tilde decomposition \cite{wootters_1998_entanglement}. In terms of the tilde orthogonal basis $\left|y_{k}\right\rangle$, the minimum decomposition of $\rho$ can be expressed directly by
\begin{equation}
\label{eqn: 53}
\rho=p_{1}\left|y_{1}\right\rangle\left\langle y_{1}\left|-\sum_{k=2}^{4} p_{k}\right| y_{k}\right\rangle\left\langle y_{k}\right|, 
\end{equation}
where $p_{k}$ is the weight corresponding to the states $\left|y_{k}\right\rangle$. Under this minimum decomposition, the concurrence of $\rho$ turns out to be the summation of the concurrence of the component pure state $Y_{k}=\left|y_{k}\right\rangle\left\langle y_{k}\right|$ as
\begin{equation}
\label{eqn: 54}
E_{c}(\rho)=p_{1} E_{c}\left(Y_{1}\right)-\sum_{k=2}^{4} p_{k} E_{c}\left(Y_{k}\right). 
\end{equation}
For the convenience of expression, we define the new states $\left|z_{1}\right\rangle=\sqrt{p}\left|y_{1}\right\rangle$ and $\left|z_{k}\right\rangle=i \sqrt{p_{k}}\left|y_{k}\right\rangle, k=2,3,4$, to rewrite \eqref{eqn: 53} as $\rho=\sum_{k}\left|z_{k}\right\rangle\left\langle z_{k}\right|$ and \eqref{eqn: 54} as
\begin{equation}
\label{eqn: 55}
E_{c}(\rho)=\sum_{k} E_{c}\left(\left|z_{k}\right\rangle\left\langle z_{k}\right|\right)=\sum_{k} E_{c}\left(Z_{k}\right).
\end{equation}
Substituting \eqref{eqn: 55} into \eqref{eqn: 52} and using the definition of concurrence for pure states given by \eqref{eqn: 24a}, we obtain
\begin{align}
\label{eqn: 56}
\begin{aligned}
    V_{c}(\rho) & =\mathcal{N}-\sum_{k} E_{c}\left(Z_{k}\right) \\
    & =\mathcal{N}-\sum_{k} \sqrt{2\left(1-\operatorname{Tr}\left(\left(Z_{k}\right)_{M}^{2}\right)\right)}.
\end{aligned}
\end{align}
In the following, the Lyapunov control law $u_{k}(\rho)$ will be derived from \eqref{eqn: 56} in terms of $\rho$ 's component pure state $Y_{k}$ to achieve the control goal $\dot{V}_{c}(\rho)<0, \forall \rho \neq \rho_{e q}$ and $\dot{V}_{c}\left(\rho_{e q}\right)=0$.

\subsection{Control Law Design}
According to \eqref{eqn: 56}, the first-order time derivative of $V_{c}(\rho)$ is
\begin{align}
\label{eqn: 57}
    \dot{V}_{c}(\rho)=-2 i \sum_{j}\left[E_{c}\left(Z_{j}\right)\right]^{-1} \operatorname{Tr}\left(\left(Z_{j}\right)_{M} \cdot\left(H Z_{j}-Z_{j} H\right)_{M}\right).
\end{align}
With the Hamiltonian $H$ expressed under the interaction picture \eqref{eqn: 6}, $\dot{V}_{c}(\rho)$ can be further simplified to
\begin{align}
\label{eqn: 58}
    \begin{aligned}
        \dot{V}_{c}(\rho)=-2 i \sum_{k} u_{k} \sum_{j}\left[E_{c}\left(Z_{j}\right)\right]^{-1}\left(\operatorname { T r } \left(\left(Z_{j}\right)_{M}\right.\right. \\
\left.\left.\cdot\left(A_{k} Z_{j}-Z_{j} A_{k}\right)_{M}\right)\right) .
    \end{aligned}
\end{align}
By a similar way taken by propostion \ref{prop: 1}, we can show that $\operatorname{Tr}\left(\left(Z_{j}\right)_{M} \cdot\left(A_{k} Z_{j}-Z_{j} A_{k}\right)_{M}\right)$ is a pure imaginary number. Thus the following quantity appears to be real-valued:
\begin{equation}
\label{eqn: 59}
x_{k}=i \sum_{j}\left[E_{c}\left(Z_{j}\right)\right]^{-1} \operatorname{Tr}\left(\left(Z_{j}\right)_{M} \cdot\left(A_{k} Z_{j}-Z_{j} A_{k}\right)_{M}\right). 
\end{equation}
Like the case of pure-state control, a real function $h_{k}\left(x_{k}\right)$ of $x_{k}$ is introduced to satisfy the conditions $h_{k}\left(x_{k}\right) \cdot x_{k} \geq 0$, and $h_{k}\left(x_{k}\right)=0$, if and only if $x_{k}=0$. The feedback signal for the mixed-state Lyapunov control then can be constructed as
\begin{equation}
\label{eqn: 60}
u_{k}=r_{k} h_{k}\left(x_{k}\right), r_{k}>0. 
\end{equation}
Substitution of \eqref{eqn: 60} into \eqref{eqn: 58} yields the desired goal of Lyapunov control
\begin{equation}
\label{eqn: 61}
\dot{V}_{c}(\rho)=-2 \sum_{k} r_{k} h_{k}\left(x_{k}\right) \cdot x_{k} \leq 0.
\end{equation}
Therefore, the control law \eqref{eqn: 60} ensures that the $\operatorname{LEF} V_{c}(\rho)$ is decreasing and meanwhile the entanglement measure $E_{c}(\rho)$ is increasing due to the relation $V_{c}(\rho)=\mathcal{N}-E_{c}(\rho)$.

\subsection{Asymptotic Stability}
The mixed state $\rho$ controlled by \eqref{eqn: 60} converges to the invariant set characterized by $\dot{V}_{c}(\rho)=0$, and from \eqref{eqn: 61} the only solution is $x_{k}=0$ because of $h_{k}\left(x_{k}\right)=0$, iff $x_{k}=0$. For arbitrary $x_{k} \neq 0$, we have $\dot{V}_{c}(\rho)<0$. With $x_{k}=0$, the control law \eqref{eqn: 60} then gives $u_{k}=0$, which in turn yields $\dot{\rho}=0$ from \eqref{eqn: 6}. Hence, the invariant set contains only the equilibrium states $\rho_{e q}$ of the von Neumann equation \eqref{eqn: 6}, which implies that the mixed state $\rho$ controlled by \eqref{eqn: 60} converges asymptotically to the equilibrium state $\rho_{e q}$. According to the properties $\dot{V}_{c}(\rho)<0$ and $\quad \dot{E}_{c}(\rho)>0, \forall \rho \neq \rho_{e q}$, and $\dot{V}_{c}\left(\rho_{e q}\right)=\dot{E}_{c}\left(\rho_{e q}\right)=0$, it appears that the equilibrium state $\rho_{e q}$ is the state that minimizes the $\operatorname{LEF} V_{c}(\rho)$ and meanwhile maximizes the entanglement measure $E_{c}(\rho)$, i.e., $\rho_{e q}=\rho_{\text{MEMS}}$. Of significance is that the value of $E_{c}\left(\rho_{e q}\right)$ automatically gives the maximum entanglement measure $E_{c}^{*}$, and we do not need to specify it in advance.

\subsection{Numerical Verification}
The Lyapunov entanglement control \eqref{eqn: 60} with $h_{k}\left(x_{k}\right)=$ $x_{k}$ is employed to obtain the MEMS. The feedback signal $x_{k}$ defined by \eqref{eqn: 59} is generated by the von Neumann equation \eqref{eqn: 6} with the process of tilde decomposition. The internal Hamiltonian is chosen as $H_{0}=\sigma_{z} \otimes \sigma_{z}$ and the control Hamiltonian $H_{k}$ is constructed in the form of
\begin{align}
    \begin{aligned}
    & H_{1}=\sigma_{z} \otimes \sigma_{x}, \quad H_{2}=\sigma_{z} \otimes \sigma_{y}, \quad H_{3}=\sigma_{y} \otimes \sigma_{z}, \\
    & H_{4}=\sigma_{x} \otimes \sigma_{z}, \quad H_{5}=\sigma_{y} \otimes \sigma_{y}, \quad H_{6}=\sigma_{y} \otimes \sigma_{x},
    \end{aligned}
\end{align}
where we note that the number of $H_{k}$ must be at least six to cover the entire range of state transfer. With the specified $H_{k}$, the time evolution of the density matrix under the interaction
picture is described by \eqref{eqn: 6} as $i \dot{\rho}(t)=\left[\sum_{k=1}^{6} u_{k} A_{k}, \rho(t)\right]$, where the control signal $u_{k}$ is determined by \eqref{eqn: 60} with gain $r_{k}=5$.

\begin{figure}[htbp]
\centering
\includegraphics[width=\columnwidth]{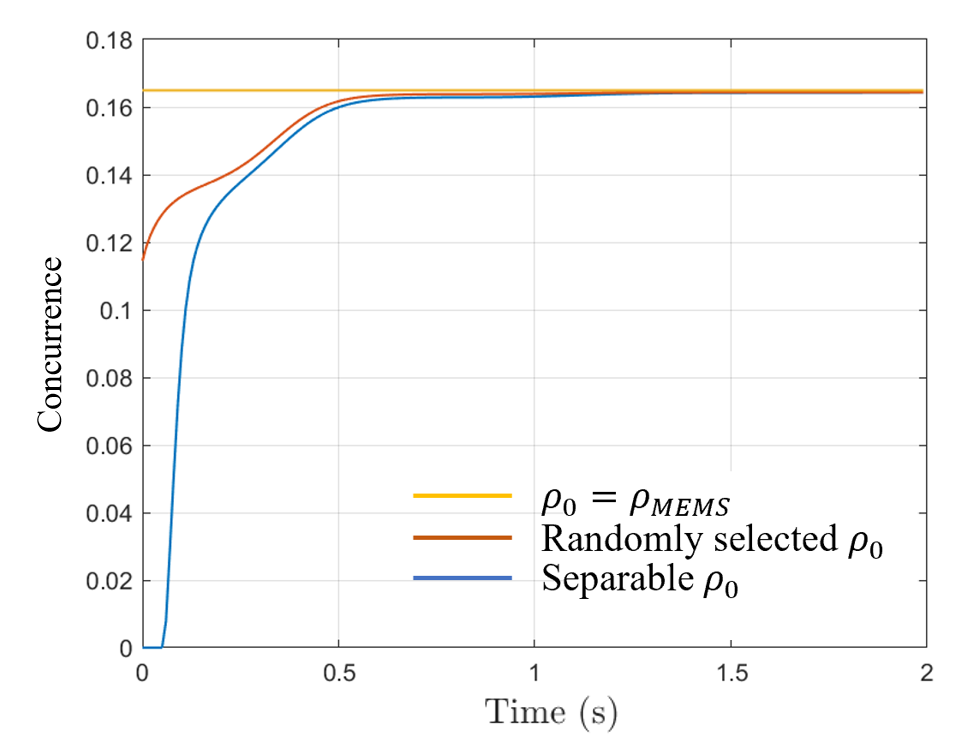}
\caption{The time responses of the mixed-state concurrence $E_{c}(\rho(t))$ under the Lyapunov entanglement control law \eqref{eqn: 60} for three initial states: $\rho_{0}=\rho_{\text{MEMS}}$ (upper flat curve), separable $\rho_{0}$ (lower curve), and a randomly selected $\rho_{0}$ between the separable $\rho_{0}$ and $\rho_{\text{MEMS}}$ (middle curve).}
\label{fig: 5}
\end{figure}
Fig. \ref{fig: 5} shows the time responses of the concurrence $E_{c}(\rho)$ by the proposed Lyapunov entanglement control for three initial states with the same spectrum $\left\{\lambda_{1}, \lambda_{2}, \lambda_{3}, \lambda_{4}\right\}=$ $\{0.4932,0.3485,0.1301,0.0282\}$. It can be seen that the time responses of the concurrence $E_{c}(\rho(t))$ all converge to the same steady state $\rho_{s s}$ with $E_{c}\left(\rho_{s s}\right)=0.1648$, which is consistent with the theoretical value $E_{c}^{*}$ as given by \eqref{eqn: 51} with the specified spectrum $\lambda_{i}$. The three curves in Fig. \ref{fig: 5} correspond to three representative initial states $\rho_{0}$. The lower curve starting from $E_{c}\left(\rho_{0}\right)=0$ is generated by a separable initial state $\rho_{0}$. The upper flat curve achieving a constant concurrence at $E_{c}^{*}$ is generated by an initial state identical to $\rho_{M E M S}$ given by \eqref{eqn: 50}, and the middle curve is generated by a randomly selected initial state between the separable state $\rho_{0}$ and $\rho_{\text {MEMS }}$.

The proposed entanglement control law \eqref{eqn: 60} is capable of searching for the MEMS from the set of the density matrices that share the same spectrum specified by $\left\{\lambda_{i}\right\}$. When the specified spectrum changes, the MEMS obtained by the control law \eqref{eqn: 60} changes accordingly. Table \ref{tab: 2} compares the steady-state concurrence $E_{c}\left(\rho_{s s}\right)$ with the theoretical value $E_{c}^{*}$ given by \eqref{eqn: 51} for ten sets of spectrum. For each spectrum, Table \ref{tab: 2} lists the steady-state values of $p_{i}$ and $E_{c}\left(Y_{i}\right)$, from which the steady-state concurrence can be computed by $E_{c}\left(\rho_{s s}\right)=p_{1} E_{c}\left(Y_{1}\right)-$ $p_{2} E_{c}\left(Y_{2}\right)-p_{3} E_{c}\left(Y_{3}\right)-p_{4} E_{c}\left(Y_{4}\right)$ as given by \eqref{eqn: 54}. The last column in Tab. \ref{tab: 2} compares the computed $E_{c}\left(\rho_{s s}\right)$ with the theoretical vaue $E_{c}^{*}$ given by \eqref{eqn: 51}.

\begin{table}[htbp]
\caption{The comparison of the steady-state concurrence $E_{c}\left(\rho_{s s}\right)$ with the theoretical value $E_{c}^{*}$ for ten sets of spectrum.}
    \label{tab: 2}
    \centering
        \includegraphics[width=\columnwidth]{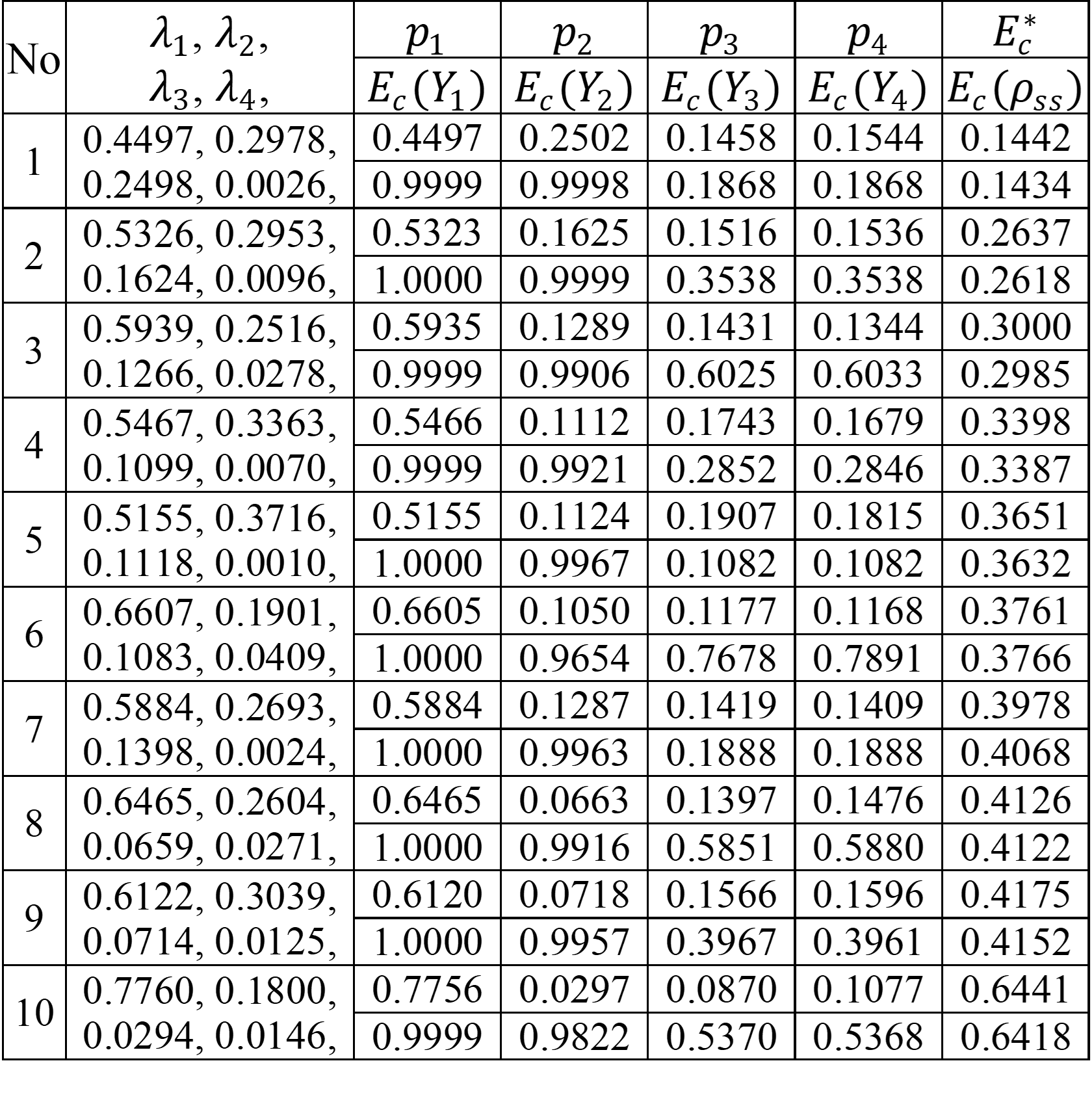}
\end{table}
If we ignore the small deviations caused by the numerical truncation errors, we find that Table \ref{tab: 2} reveals some significant regularities. For a given spectrum $\left\{\lambda_{1}, \lambda_{2}, \lambda_{3}, \lambda_{4}\right\}$ with decreasing order, the steady-state value of $E_{c}\left(Y_{k}\right)$ given by Table \ref{tab: 2} demonstrates the following regularity
\begin{align}
\label{eqn: 63}
\begin{aligned}
    & E_{c}\left(Y_{1}\right)=E_{c}\left(Y_{2}\right)=1 \\
    & E_{c}\left(Y_{3}\right)=E_{c}\left(Y_{4}\right)=2 \sqrt{\lambda_{2} \lambda_{4}} /\left(\lambda_{2}+\lambda_{4}\right).
\end{aligned} 
\end{align}
Meanwhile, the steady-state weight $p_{k}$ of $E_{c}\left(Y_{k}\right)$ shows the following regularity
\begin{equation}
\label{eqn: 64}
p_{1}=\lambda_{1}, p_{2}=\lambda_{3}, p_{3}=p_{4}=\left(\lambda_{2}+\lambda_{4}\right) / 2 .
\end{equation}
The combination of \eqref{eqn: 63} and \eqref{eqn: 64} gives an error-free prediction of $E_{c}\left(\rho_{s s}\right)$ as
\begin{align}
\label{eqn: 65}
\begin{aligned}
    E_{c}\left(\rho_{ss}\right) & =p_{1} E_{c}\left(Y_{1}\right)-p_{2} E_{c}\left(Y_{2}\right)-p_{3} E_{c}\left(Y_{3}\right)-p_{4} E_{c}\left(Y_{4}\right) \\
& =\lambda_{1}-\lambda_{3}-2 \sqrt{\lambda_{2} \lambda_{4}}
\end{aligned}
\end{align}
which recovers the theoretical result \eqref{eqn: 51}.

By comparing with the analytical solution, the above numerical results confirm that the proposed Lyapunov entanglement control can precisely generate the MEMS. More importantly, we do not need to specify in advance the MEMS to be generated during the control process. It is because of this property that we can discover more different forms of MEMS not belonging to the known class generated by the kernel mixed state \eqref{eqn: 50}.

The tilde decomposition of the MEMS in the class generated by \eqref{eqn: 50} possesses the properties expressed by \eqref{eqn: 63} and \eqref{eqn: 64}. However, there are many MEMS outside this class. For example, considering the following MEMS
\begin{align}
\label{eqn: 66}
\begin{aligned}
    \rho & =\lambda_{1}\left|\beta_{00}\right\rangle\left\langle\beta_{00}\left|+\sqrt{\lambda_{2} \lambda_{4}}\right| \beta_{10}\right\rangle\left\langle\beta_{10}\right|  \\
& +\lambda_{3}\left|\beta_{01}\right\rangle\left\langle\beta_{01}\left|+\sqrt{\lambda_{2} \lambda_{4}}\right| \beta_{11}\right\rangle\left\langle\beta_{11}\right|,
\end{aligned}
\end{align}
we find that its tilde decomposition has the property $E_{c}\left(Y_{1}\right)=$ $E_{c}\left(Y_{2}\right)=E_{c}\left(Y_{3}\right)=E_{c}\left(Y_{4}\right)=1$, which is different from the pattern specified by \eqref{eqn: 63}. It is clear that the MEMS given by \eqref{eqn: 66} does not belong to the class generated by \eqref{eqn: 50}; however, it still achieves the maximum concurrence $E_{c}^{*}$ given by \eqref{eqn: 51}.

Table \ref{tab: 3} lists the MEMS generated by the Lyapunov entanglement control law, which otherwise can not be obtained by applying any local unitary transformation to \eqref{eqn: 50}. It can be checked that the steady-state values of $p_{i}$ and $E_{c}\left(Y_{\mathrm{i}}\right)$ listed in Tab. \ref{tab: 3} do not have the regularities expressed by \eqref{eqn: 63} and \eqref{eqn: 64}, indicating that the class of MEMS in Table \ref{tab: 3} is different from the class covered by Table \ref{tab: 2}. Nevertheless, we note that although
the MEMSs in Tables \ref{tab: 2} and \ref{tab: 3} belong to different classes, they all attain the maximum concurrence $E_{c}^{*}$ within the numerical accuracy. The last column in Table \ref{tab: 3} compares the computed $E_{c}\left(\rho_{s s}\right)=p_{1} E_{c}\left(Y_{1}\right)-\sum_{k=2}^{4} p_{k} E_{c}\left(Y_{k}\right)$ with $E_{c}^{*}$.
\begin{table}[htbp]
\caption{Several MEMSs not belonging to the class covered by \eqref{eqn: 50}.}
    \label{tab: 3}
    \centering
        \includegraphics[width=\columnwidth]{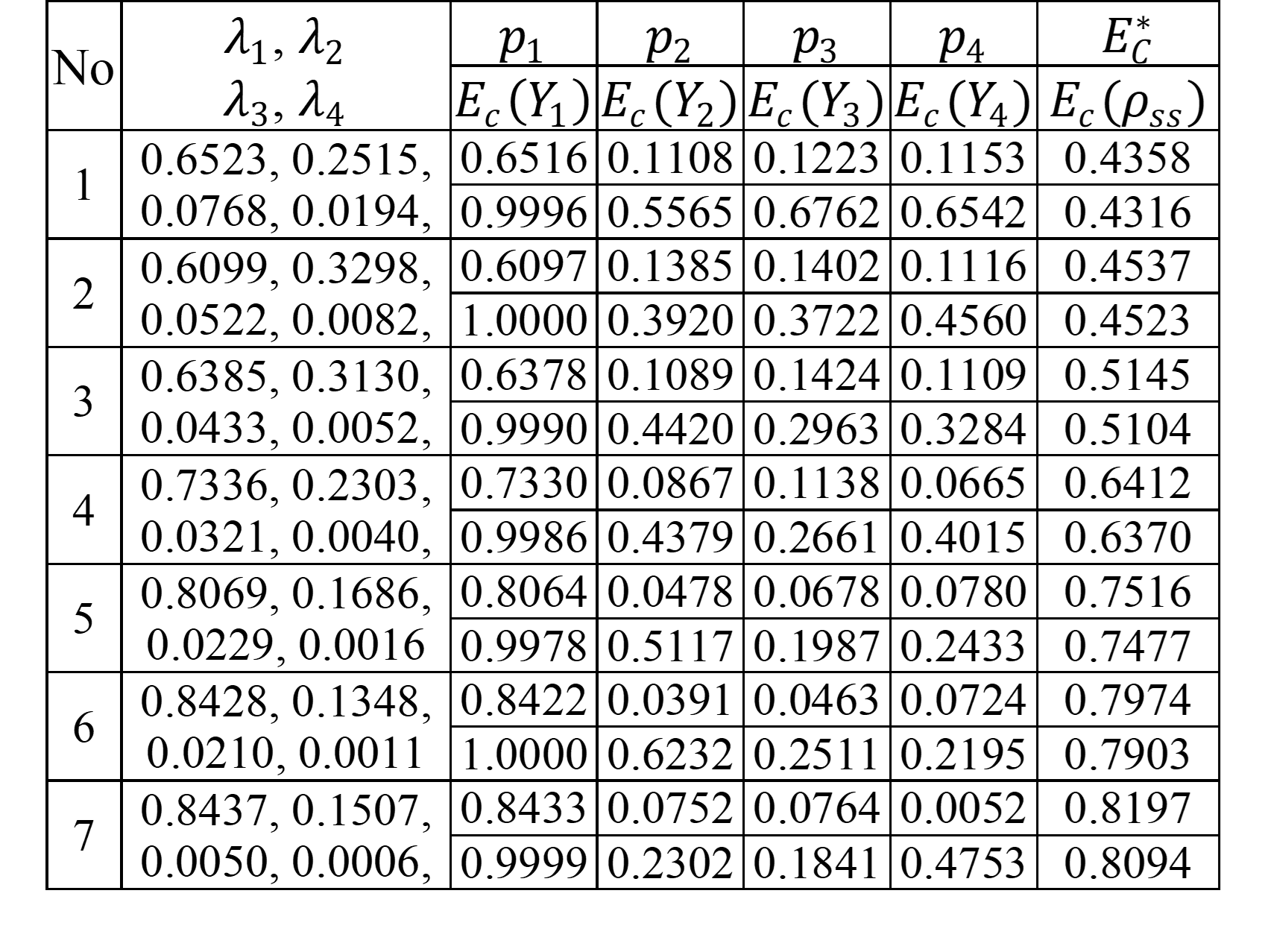}
\end{table}
Regarding the maximum entanglement control for mixed states, Tables \ref{tab: 2} and \ref{tab: 3} present two of the major results of this paper. The former shows that the MEMS obtained by the proposed method is completely consistent with the analytical solution mentioned in the literature, and the latter shows that our control method can also be used to generate new forms of MEMS.

\section{Lyapunov entanglement control for multipartite systems}
\label{sec: Lyapunov_design_multi}
The same entanglement control strategy that has been used for the pure state and the mixed state of bipartite systems can be applied to multipartite systems. For multipartite systems, the main challenge lies not in the formulation of the Lyapunov entanglement control, but in that the currently available multipartite entanglement measures can only determine a lower bound, but not the exact entanglement of a multipartite state. Furthermore, the MES of multipartite systems is not unique, because using different entanglement measures may result in different MES, such as W state or GHZ state, between which there is no unitary local transformation.
In this section, two entanglement measures for multipartite systems, i.e., generalized concurrence and genuine multipartite entanglement, are employed in the Lyapunov entanglement control to generate the multipartite MES. 
\subsection{Two Entanglement measures for multipartite states}
Generalized concurrence \cite{li_2009_a} provides a lower bound of the degree of multipartite entanglement. Let $\mathcal{H}_{j}$ denote a Hilbert space with dimension $d$, whose basis is given by $\left|k_{j}\right\rangle$, $k_{j}=1,2 \ldots d$. A $N$-partite pure state in the space of $\mathcal{H}_{1} \otimes$ $\mathcal{H}_{2} \otimes \cdots \otimes \mathcal{H}_{N}$ is represented by
\begin{equation}
\label{eqn: 67}
|\psi\rangle=\sum_{k_{1}=1}^{d} \sum_{k_{2}=1}^{d} \cdots \sum_{k_{N}=1}^{d} a_{k_{1}, k_{2}, \cdots k_{N}}\left|k_{1}, k_{2}, \cdots k_{N}\right\rangle .
\end{equation}
Let $\alpha$ and $\alpha^{\prime}$ (resp. $\beta$ and $\beta^{\prime}$ ) be the subsets of the index set $\left\{k_{1}, k_{2}, \cdots k_{N}\right\}$, which are associated with the same Hilbert spaces but with different summing indices, so that $\{\alpha, \beta\}=$ $\left\{\alpha^{\prime}, \beta^{\prime}\right\}=\left\{k_{1}, k_{2}, \cdots, k_{N}\right\}$. Then the generalized concurrence of $|\psi\rangle$ is given by
\begin{equation}
\label{eqn: 68}
E_{\text{GC}}=\sqrt{\frac{d}{2 m(d-1)} \sum_{p} \sum_{\left\{\alpha, \alpha^{\prime}, \beta, \beta^{\prime}\right\}}^{d}\left|a_{\alpha \beta} a_{\alpha^{\prime} \beta^{\prime}}-a_{\alpha \beta^{\prime}} a_{\alpha^{\prime} \beta}\right|^{2}} 
\end{equation}
where $m=2^{N-1}-1$ and the outer summation is over all possible combinations of the two subsets $\alpha$ and $\beta$. For a $N$-bit state $|\psi\rangle$, we have $d=2$, for which \eqref{eqn: 68} can be simplified to
\begin{equation}
\label{eqn: 69}
E_{\text{GC}}(\rho)=\sqrt{\frac{1}{2^{N-1}-1}\left(N-\sum_{j=1}^{N} \operatorname{Tr}\left(\rho_{j}^{2}\right)\right)} 
\end{equation}
where $\rho_{j}=\operatorname{Tr}_{j}(\rho)$ is the reduced matrix of $\rho=|\psi\rangle\langle\psi|$ obtained by taking the partial trace for all the subsystems except the $j$ th particle. When $N=2$, \eqref{eqn: 69} reduces to \eqref{eqn: 24a} for bipartite pure states by noting $\rho_{1}=\rho_{2}=\rho_{M}$. The LEF for $E_{\text{GC}}(\rho)$ can be chosen as
\begin{equation}
\label{eqn: 70}
V_{\text{GC}}(\rho)=\mathcal{N}-E_{\text{GC}}(\rho)
\end{equation}
where $\mathcal{N}$ is a trivial constant. The multipartite Lyapunov control can be derived from the condition $\dot{V}_{\text{GC}}=-\dot{E}_{\text{GC}} \leq 0$ to drive $\rho=|\psi\rangle\langle\psi|$ from an arbitrary state $\rho_{0}$ to the steady state $\rho_{\text{eq}}$, at which $E_{\text{GC}}$ achieves its maximum.

The same control strategy can be applied to other qualified entanglement measures for multipartite systems, such as genuine-multipartite-entanglement (GME) concurrence \cite{ma_2011_measure, chen_2012_improved}, which searches for the minimum bipartition of the system. A pure $N$-partite state $|\psi\rangle \in \mathcal{H}_{1} \otimes \mathcal{H}_{2} \otimes \cdots \otimes \mathcal{H}_{N}$ is said to be biseparable if it can be written as $|\psi\rangle=\left|\psi_{A}\right\rangle \otimes\left|\psi_{B}\right\rangle$, where $\quad\left|\psi_{A}\right\rangle \in \mathcal{H}_{A}=\mathcal{H}_{j_{1}} \otimes \ldots \otimes \mathcal{H}_{j_{k}} \quad$ and $\quad\left|\psi_{B}\right\rangle \in \mathcal{H}_{B}=$ $\mathcal{H}_{j_{k+1}} \otimes \ldots \otimes \mathcal{H}_{j_{N}}$; otherwise, it is said to be genuine $N$ partite entangled. Supposing $\left\{\gamma_{k} \mid \gamma_{k}^{\prime}\right\}=\left\{j_{1}, j_{2}, \ldots j_{k} \mid j_{k+1}, \cdots j_{N}\right\}$ is a bi-partition of the index set $\{1,2, \cdots, n\}$, GME concurrence searches for the particular bi-partition of the system to minimize the concurrence:
\begin{equation}
\label{eqn: 71}
E_{\text{GME}}(\rho)=\min _{\left\{\gamma_{k} \mid \gamma_{k}^{\prime}\right\}} \sqrt{2\left(1-\operatorname{Tr}\left(\rho_{\gamma_{k}}^{2}\right)\right)} 
\end{equation}
where $\rho_{\gamma_{k}}$ is the reduced density matrix obtained by taking the partial trace of $\rho$ over the subsystem indexed by $\gamma_{k}^{\prime}$ and the minimization is over all possible bipartitions $\left\{\gamma_{k} \mid \gamma_{k}^{\prime}\right\}$. Similarly, the LEF for $E_{\text{GME}}(\rho)$ can be chosen as
\begin{equation}
\label{eqn: 72}
V_{\text{GME}}(\rho)=\mathcal{N}-E_{\text{GME}}(\rho) .
\end{equation}
Lyapunov control aims to drive $E_{\text{GME}}(\rho)$ to its maximum and to compare with the theoretical maximum achieved by the $|\text{GHZ}\rangle$ state.

\subsection{Designing Lyapunov entanglement control laws}
Once a qualified LEF is chosen for multipartite states, the design of Lyapunov entanglement control is the same as that of bipartite states without additional difficulty. We start with the LEF of the general concurrence $V_{G C}(\rho)$ given by \eqref{eqn: 70}, whose first-order time derivative can be expressed as
\begin{align}
\label{eqn: 73}
\begin{aligned}
\dot{V}_{\text{GC}} & =\frac{E_{\text{GC}}^{-1}}{2^{N-1}-1} \sum_{j=1}^{N} \operatorname{Tr}\left(\rho_{j} \dot{\rho}_{j}\right) \\
& =-i \frac{E_{\text{GC}}^{-1}}{2^{N-1}-1} \sum_{j=1}^{N} \operatorname{Tr}\left(\rho_{j} \cdot(H \rho-\rho H)_{j}\right. 
\end{aligned}
\end{align}
The system Hamiltonian $H$ under the interaction picture is given by \eqref{eqn: 6} with which the relation of $\dot{V}_{G C}$ to the control field $u_{k}$ can be derived as
\begin{equation}
\label{eqn: 74}
\dot{V}_{\text{GC}}=-i \frac{E_{\text{GC}}^{-1}}{2^{N-1}-1} \sum_{k=1}^{r} u_{k} \sum_{j=1}^{N} \operatorname{Tr}\left(\rho_{j}\left(A_{k} \rho-\rho A_{k}\right)_{j}\right) . 
\end{equation}
No matter how many subsystems the system has, the Hermitian property of the density matrix $\rho$ and its reduced form $\rho_{j}$ does not change. As a result, we can show by the same way used in Proposition \ref{prop: 1} that $\sum_{j} \operatorname{Tr}\left(\rho_{j} \cdot\left(A_{k} \rho-\rho A_{k}\right)_{j}\right)$ is an imaginary number. In terms of the real-valued variable,
\begin{equation}
\label{eqn: 75}
x_{k}=i \cdot \sum_{j=1}^{N} \operatorname{Tr}\left(\rho_{j}\left(A_{k} \rho-\rho A_{k}\right)_{j}\right),
\end{equation}
the desired Lyapunov control law now can be contructed as $u_{k}=r_{k} h_{k}\left(x_{k}\right)$, where $h_{k}\left(x_{k}\right)$ satisfies the relation \eqref{eqn: 35} and $r_{k}$ is a positive control gain. With $u_{k}=r_{k} h_{k}\left(x_{k}\right)$, the time derivative of $V_{\text{GC}}$ becomes
\begin{equation}
\label{eqn: 76}
\dot{V}_{\text{GC}}=-\frac{E_{\text{GC}}^{-1}}{2^{N-1}-1} \sum_{k=1}^{r} r_{k} x_{k} h_{k}\left(x_{k}\right) \leq 0, \forall t \geq 0 .
\end{equation}
It can be shown from the property of $h_{k}\left(x_{k}\right)$ that $\dot{V}_{\text{GC}}=\dot{E}_{\text{GC}}=0$ occurs only at the equilibrium state $\rho_{e q}$, and $\dot{V}_{\text{GC}}=-\dot{E}_{\text{GC}}<$ $0, \forall \rho \neq \rho_{e q}$. Therefore, the proposed control law drives $\rho$ to the equilibrium state $\rho_{\text{eq}}$, where $V_{\text{GC}}$ reaches its minimum and $E_{\text{GC}}$ reaches its maximum, i.e., $\rho_{e q}$ is the MES of $E_{\text{GC}}$.

Next, we consider the entanglement control of GME-concurrence defined by \eqref{eqn: 71}. Let $\gamma_{m}$ be the partition $\gamma_{k}$ that attains the minimum in \eqref{eqn: 71} at time $t$, then the first-order time derivative of $V_{\text{GME}}$ at time $t$ can be expressed as
\begin{equation}
\label{eqn: 77}
\dot{V}_{\text{GME}}=-2 i E_{\text{GME}}^{-1} \cdot \sum_{k=1}^{r} u_{k} \operatorname{Tr}\left(\rho_{\gamma_{m}} \cdot\left(A_{k} \rho-\rho A_{k}\right)_{\gamma_{m}}\right) .
\end{equation}
To ensure $\dot{V}_{\text{GME}} \leq 0$, the control law $u_{k}=r_{k} h_{k}\left(x_{k}\right)$ is applied again with the real-valued feedback signal $x_{k}$ defined by
\begin{equation}
\label{eqn: 78}
x_{k}=i \cdot \operatorname{Tr}\left(\rho_{\gamma_{m}} \cdot\left(A_{k} \rho-\rho A_{k}\right)_{\gamma_{m}}\right).
\end{equation}
This control law yields
\begin{align}
   \dot{V}_{\text{GME}}=-\dot{E}_{\text{GME}}=-2 \sum_{k=1}^{r} r_{k} x_{k} h_{k}\left(x_{k}\right) \leq 0, \forall t \geq 0, 
\end{align}
which drives $E_{\text{GME}}$ to its maximum at the equilibrium state $\rho_{\text{eq}}$.

\subsection{Numerical verification}
This section demonstrates the convergence of the proposed Lyapunov entanglement control towards the MES using tripartite quantum states. Firstly, the internal Hamiltonian $H_{0}$ and the control Hamiltonian $H_{k}$ adopt the same setting values as in Section \ref{sec: numerical verification of maximum entanglement control}. The LEFs for the two entanglement measures $E_{\text{GC}}$ and $E_{\text{GME}}$ with $N=3$ are given, respectively, by \eqref{eqn: 70} and \eqref{eqn: 72} as
\begin{subequations}
\renewcommand{\theequation}{\theparentequation\alph{equation}}
    \begin{equation}
        \label{eqn: 80a}
        V_{\text{GC}}  =\mathcal{N}-\sqrt{\left(1-\frac{1}{3}\sum_{k=1}^{3} \operatorname{Tr}\left(\rho_{k}^{2}\right)\right) },
    \end{equation}
    \begin{equation}
        \label{eqn: 80b}
        V_{\text{GME}}  =\mathcal{N}-\min _{k \in\{1,2,3\}} \sqrt{2\left(1-\operatorname{Tr}\left(\rho_{k}^{2}\right)\right)} .
    \end{equation}
\end{subequations}
The same Lyapunov control law $u_{k}=r_{k} h_{k}\left(x_{k}\right)=5 x_{k}$ is applied to the two entanglement measures, where the feedback signal $x_{k}$ for $V_{\text{GC}}$ and $V_{\text{GME}}$ is calculated from \eqref{eqn: 75} and \eqref{eqn: 78}, respectively. Two initial states are tested in the numerical demonstration: one is the separable state $\rho_{0}=|000\rangle$, and the other is a randomly selected inseparable state.

\begin{figure}[htbp]
\centering
\includegraphics[width=\columnwidth]{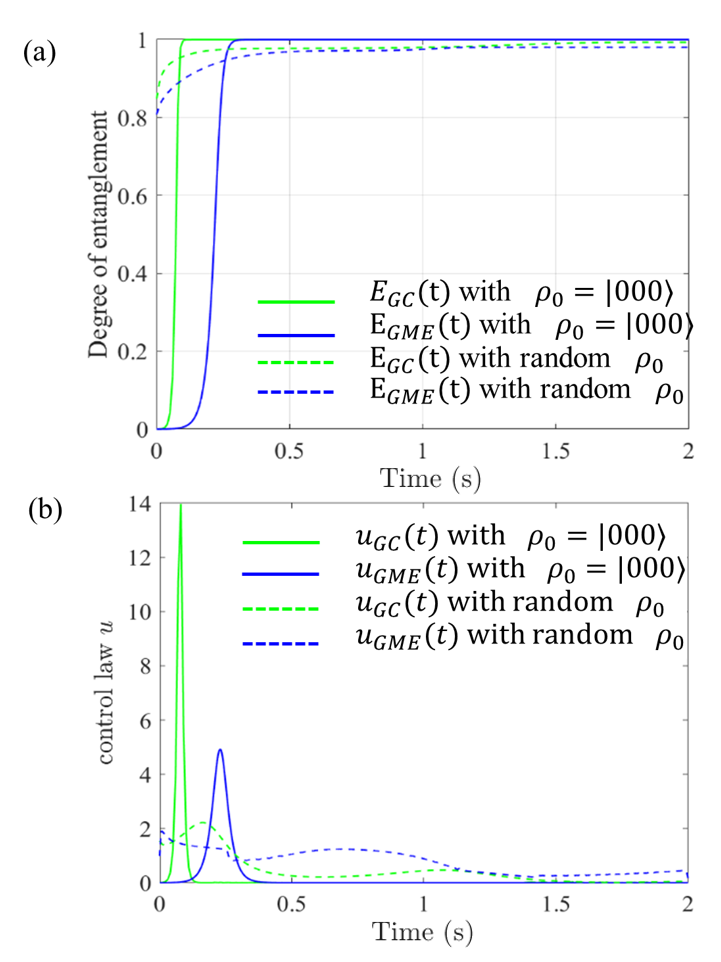}
\caption{The time responses of $E_{\text{GC}}(t)$ and $E_{\text{GME}}(t)$ and their control magnitudes $u_{\text{GC}}(t)$ and $u_{\text{GME}}(t)$ starting from two initial states. Both $E_{\text{GC}}(t)$ and $E_{\text{GME}}(t)$ converge to the theoretical maximum equal to $1$, with the former converging faster than the latter.}
\label{fig: 6}
\end{figure}

The resulting time responses of $E_{\text{GC}}(t)$ and $E_{\text{GME}}(t)$ are shown in Fig. \ref{fig: 6} (a). As expected, the proposed Lyapunov control law drives both $E_{\text{GC}}(t)$ and $E_{\text{GME}}(t)$ to their maximum, which is equal to that achieved theoretically by the $|\text{GHZ}\rangle$ state. $E_{\text{GC}}(t)$ converges faster than $E_{\text{GME}}(t)$, but consumes much more control energy, as shown in Fig. \ref{fig: 6} (b). The other noticeable observation from Fig. \ref{fig: 6} is that the Lyapunov entanglement control starting from different initial states $\rho_{0}$ may converge to different MES with different convergence speeds. The convergence to the MESs from the randomly selected $\rho_{0}$ is slower than that from $\rho_{0}=|000\rangle$. Collecting all the MESs generated by the Lyapunov entanglement control from different initial states forms a class of MES whose entanglement measure is equal to that of $|\text{GHZ}\rangle$. Although the degree of entanglement of the obtained MESs is the same as that of $|\text{GHZ}\rangle$, it does not mean that we can apply the LOCC operation to convert these MESs into the $|\text{GHZ}\rangle$ state. This is the main difference from the bipartite entanglement control, for which all the obtained MESs are equivalent to the Bell state by LOCC operation.

\section{CONCLUSION AND FUTURE WORK}
This paper proposed a control design approach for generating MES by constructing the Lyapunov function from an entanglement measure $E(\rho)$. The proposed control design enables the generation of MESs through control design without requiring prior knowledge of their forms. The number of entangled subsystems is not a limiting factor as long as the entanglement measure $E(\rho)$ accurately describes the degree of entanglement of the quantum state $\rho$. While demonstrating the feasibility of the proposed method, further investigation is necessary to understand the general structure of generated MES, especially for multipartite systems, through analyzing LaSalle's invariant set. Additionally, the Lyapunov control law can be refined to increase the convergence speed of MES. Furthermore, applying this method to open quantum systems can help elucidate the influence of external interactions on the preparation of MES.

Moving forward, we plan to deepen our understanding of multipartite entanglement measures and enhance the control strategy's robustness against perturbations and noise. We also intend to evaluate alternative measures and examine the physical realizability of the control protocol in experimental setups. These efforts can pave the way for further advancements in quantum control techniques, bringing us closer to realizing practical quantum information processing tasks and quantum communication protocols.

\bibliography{apssamp}

\end{document}